\documentclass[twocolumn,twocolappendix]{aastex6}
\newcommand{\colwidth}{\linewidth}

\usepackage{graphicx}	
\usepackage{amsmath}	
\usepackage{amssymb}	
\usepackage{bm}		    

\usepackage{mathrsfs}
\usepackage{microtype}

\usepackage{savesym}
\savesymbol{tablenum}
\usepackage{siunitx}
\restoresymbol{SIX}{tablenum}
\newcommand\abs[1]{\left|#1\right|}

\usepackage{natbib}     

\usepackage{lineno}

\usepackage{tikz}
\usetikzlibrary{arrows,positioning,shapes,decorations.markings}

\newif\ifref
\reffalse
\newcommand{\mb}[1]{\ifref\boldmath\textbf{#1}\unboldmath\else #1\fi}

\shorttitle{Stellar Parameters in an Instant with Machine Learning}
\shortauthors{Bellinger \& Angelou et al.}

\begin{document}

\title{Fundamental Parameters of Main-Sequence Stars in an Instant with Machine Learning}

\author{Earl P. Bellinger\altaffilmark{1,2,3}, George C. Angelou\altaffilmark{1,2}, Saskia Hekker\altaffilmark{1,2}, Sarbani Basu\altaffilmark{4}, Warrick H. Ball\altaffilmark{5}, and Elisabeth Guggenberger\altaffilmark{1,2}}
\affil{\altaffilmark{1} Max-Planck-Institut f{\"u}r Sonnensystemforschung, Justus-von-Liebig-Weg 3, 37077 G{\"o}ttingen, Germany\\
\altaffilmark{2} Stellar Astrophysics Centre, Department of Physics and Astronomy, Aarhus University, Ny Munkegade 120, DK-8000 Aarhus C, Denmark \\
\altaffilmark{3} Institut f{\"u}r Informatik, Georg-August-Universit{\"a}t G{\"o}ttingen, Goldschmidtstrasse 7, 37077 G{\"o}ttingen, Germany \\
\altaffilmark{4} Department of Astronomy, Yale University, New Haven, CT 06520, USA \\
\altaffilmark{5} Institut f{\"u}r Astrophysik G{\"o}ttingen, Friedrich-Hund-Platz 1, 37077 G{\"o}ttingen, Germany}

\begin{abstract}
Owing to the remarkable photometric precision of space observatories like \emph{Kepler}, stellar and planetary systems beyond our own are now being characterized en masse for the first time. These characterizations are pivotal for endeavors such as searching for Earth-like planets and solar twins, understanding the mechanisms that govern stellar evolution, and tracing the dynamics of our Galaxy. The volume of data that is becoming available, however, brings with it the need to process this information accurately and rapidly. While existing methods can constrain \mb{fundamental stellar parameters such as ages, masses, and radii} from these observations, they require substantial computational efforts to do so. 

We develop a method based on machine learning for rapidly estimating fundamental parameters of main-sequence solar-like stars from classical and asteroseismic observations. We first demonstrate this method on a hare-and-hound exercise and then apply it to the Sun, 16 Cyg A \& B, and 34 planet-hosting candidates that have been observed by the \emph{Kepler} spacecraft. We find that our estimates and their associated uncertainties are comparable to the results of other methods, but with the additional benefit of being able to explore many more stellar parameters while using much less computation time. We furthermore use this method to present evidence for an empirical diffusion-mass relation. Our method is open source and freely available for the community to use.\footnote{The source code for all analyses and for all figures appearing in this manuscript can be found electronically at \url{https://github.com/earlbellinger/asteroseismology} \citep{earl_bellinger_2016_55400}.}
\end{abstract}


\section{Introduction}

In recent years, dedicated photometric space missions have delivered dramatic improvements to time-series observations of solar-like stars. These improvements have come not only in terms of their precision, but also in their time span and sampling, which has thus enabled direct measurement of dynamical stellar phenomena such as pulsations, binarity, and activity. Detailed measurements like these place strong constraints on models used to determine the ages, masses, and chemical compositions of these stars. This in turn facilitates a wide range of applications in astrophysics, such as testing theories of stellar evolution, characterizing extrasolar planetary systems \citep[e.g.][]{2015ApJ...799..170C, 2015MNRAS.452.2127S}, assessing galactic chemical evolution \citep[e.g.][]{2015ASSP...39..111C}, and performing ensemble studies of the Galaxy \citep[e.g.][]{2011Sci...332..213C, 2013MNRAS.429..423M, 2014ApJS..210....1C}. 

The motivation to increase photometric quality has in part been driven by the goal of measuring oscillation modes in stars that are like our Sun. Asteroseismology, the study of these oscillations, provides the opportunity to constrain the ages of stars through accurate inferences of their interior structures. However, stellar ages cannot be measured directly; instead, they depend on indirect determinations via stellar modelling. 

Traditionally, to determine the age of a star, procedures based on iterative optimization (hereinafter IO) seek the stellar model that best matches the available observations \citep{1994ApJ...427.1013B}. 
Several search strategies have been employed, including exploration through a pre-computed grid of models (i.e.\ grid-based modelling, hereinafter GBM; see \citealt{2011ApJ...730...63G, 2014ApJS..210....1C}); or \emph{in situ} optimization (hereinafter ISO) such as genetic algorithms \citep{2014ApJS..214...27M}, Markov-chain Monte Carlo \citep{2012MNRAS.427.1847B}, or the downhill simplex algorithm (\citealt{2013ApJS..208....4P}; see e.g.\ \citealt{2015MNRAS.452.2127S} for an extended discussion on the various methods of dating stars). Utilizing the detailed observations from the \emph{Kepler} and CoRoT space telescopes, these procedures have constrained the ages of several field stars to within 10\% of their main-sequence lifetimes \citep{2015MNRAS.452.2127S}. 

IO is computationally intensive in that it demands the calculation of a large number of stellar models (see \citealt{2009ApJ...699..373M} for a discussion). ISO requires that new stellar tracks are calculated for each target, as they do not know \emph{a priori} all of the combinations of stellar parameter values that the optimizer will need for its search. They furthermore converge to local minima and therefore need to be run multiple times from different starting points to attain global coverage. GBM by way of interpolation in a high-dimensional space, on the other hand, is sensitive to the resolution of each parameter and thus requires a very fine grid of models to search through \citep[see e.g.][who use more than five million models that were varied in just four initial parameters]{2010ApJ...725.2176Q}. Additional dimensions such as efficiency parameters (e.g.\ overshooting or mixing length parameters) significantly impact on the number of models needed and hence the search times for these methods. As a consequence, these approaches typically use, for example, a solar-calibrated mixing length parameter or a fixed amount of convective overshooting. Since these values in other stars are unknown, keeping them fixed therefore results in underestimations of uncertainties. This is especially important in the case of atomic diffusion, which is essential when modelling the Sun \citep[see e.g.][]{1994MNRAS.269.1137B}, but is usually disabled for stars with M/M$_\odot > 1.4$ because it leads to the unobserved consequence of a hydrogen-only surface \citep{2002A&A...390..611M}. 

These concessions have been made because the relationships connecting \mb{observations} of stars to their internal \mb{properties} are non-linear and difficult to characterize. Here we will show that through the use of machine learning, it is possible to avoid these difficulties by capturing those relations statistically and using them to construct a regression model capable of relating observations of stars to their structural, chemical, and evolutionary properties. The relationships can be learned using many fewer models than IO methods require, and can be used to process entire stellar catalogs with a cost of only seconds per star. 

To date, only about a hundred solar-like oscillators have had their frequencies resolved, allowing each of them be modelled in detail using costly methods based on IO. In the forthcoming era of TESS \citep{2015JATIS...1a4003R} and PLATO \citep{2014ExA....38..249R}, however, seismic data for many more stars will become available, and it will not be possible to dedicate large amounts of supercomputing time to every star. Furthermore, for many stars, it will only be possible to resolve \emph{global} asteroseismic quantities rather than individual frequencies. Therefore, the ability to rapidly constrain stellar parameters for large numbers of stars by means of global oscillation analysis will be paramount. 

In this work, we consider the constrained multiple-regression problem of inferring fundamental stellar \mb{parameters} from observable \mb{quantities}. We construct a random forest of decision tree regressors to learn the relationships connecting observable quantities of main-sequence (MS) stars to their zero-age main-sequence (ZAMS) histories and  current-age structural and chemical attributes. We validate our technique by inferring the parameters of simulated stars in a hare-and-hound exercise, the Sun, and the well-studied stars 16 Cyg A and B. Finally, we conclude by applying our method on a catalog of \emph{Kepler} objects-of-interest (hereinafter KOI; \citealt{2016MNRAS.456.2183D}).  

We explore various model physics by considering stellar evolutionary tracks that are varied not only in their initial mass and chemical composition, but also in their efficiency of convection, extent of convective overshooting, and strength of gravitational settling. We compare our results to the recent findings from GBM \citep{2015MNRAS.452.2127S}, ISO \citep{2015ApJ...811L..37M}, interferometry \citep{2013MNRAS.433.1262W}, and asteroseismic glitch analyses \citep{2014ApJ...790..138V} and find that we obtain similar estimates but with orders-of-magnitude speed-ups.

\section{Method} \label{sec:Method} 
We seek a multiple-regression model capable of characterizing observed stars. To obtain such a model, we build a matrix of evolutionary simulations and use machine learning to discover relationships in the \mb{stellar models} that connect observable quantities of stars to the model quantities that we wish to predict. \mb{The matrix is structured such that each column contains a different stellar quantity and each row contains a different stellar model.} We construct this matrix by extracting models along evolutionary sequences (see Appendix \ref{sec:selection} for details on the model selection process) and summarizing them to yield the same types of information as the stars being observed. Although each star (and each stellar model) may have a different number of \mb{oscillation} modes observed, it is possible to condense this information into only a few numbers by leveraging the fact that the frequencies of these modes follow a regular pattern \citep[for a review of solar-like oscillations, see][]{doi:10.1146/annurev-astro-082812-140938}. Once the machine has processed this matrix, one can feed the algorithm a catalogue of \mb{stellar observations} and use it to predict the \mb{fundamental} parameters of those stars.

The \mb{observable information obtained from models that can be} used to inform the algorithm may include, but \mb{is} not limited to, combinations of temperatures, metallicities, global oscillation information, surface gravities, luminosities, and/or radii. From these, the machine can learn how to infer stellar parameters such as ages, masses, core hydrogen and surface helium abundances. If luminosities, surface gravities, and/or radii are not supplied, then they may be predicted as well. In addition, the machine can also infer evolutionary parameters such as the initial stellar mass and initial chemical compositions as well as the mixing length parameter, overshoot coefficient, and diffusion multiplication factor needed to reproduce observations, which are explained in detail below. 

\subsection{Model Generation}
\label{sec:models}
We use the open-source 1D stellar evolution code \emph{Modules for Experiments in Stellar Astrophysics} \citep[MESA;][]{Paxton2011} to generate main-sequence stellar models from solar-like evolutionary tracks varied in initial mass M, helium Y$_0$, metallicity Z$_0$, mixing length parameter $\alpha_{\text{MLT}}$, overshoot coefficient $\alpha_{\text{ov}}$, and \mb{diffusion multiplication factor} D. \mb{The diffusion multiplication factor} serves to amplify or diminish the effects of diffusion, where a value of zero turn\mb{s} it off and a value of two double\mb{s} all velocities. The initial conditions are varied in the ranges M $\in [0.7, 1.6]$ M$_\odot$, Y$_0$ $\in [0.22, 0.34]$, Z$_0$ $\in [10^{-5}, 10^{-1}]$ (varied logarithmically), $\alpha_{\text{MLT}}$ $\in [1.5, 2.5]$, $\alpha_{\text{ov}}$ $\in [10^{-4}, 1]$ (varied logarithmically), and D $\in [10^{-6}, 10^2]$ (varied logarithmically). We put a cut-off of 10$^{-3}$ and 10$^{-5}$ on $\alpha_{\text{ov}}$ and D, respectively, below which we consider them to be zero and \mb{disable them}. The initial parameters of each track are chosen in a quasi-random fashion so as to populate the initial-condition hyperspace as homogeneously and rapidly as possible (shown in Figure \ref{fig:inputs}; see Appendix \ref{sec:grid} for more details). 

\begin{figure*}
    \centering
    \includegraphics[width=\linewidth,keepaspectratio]{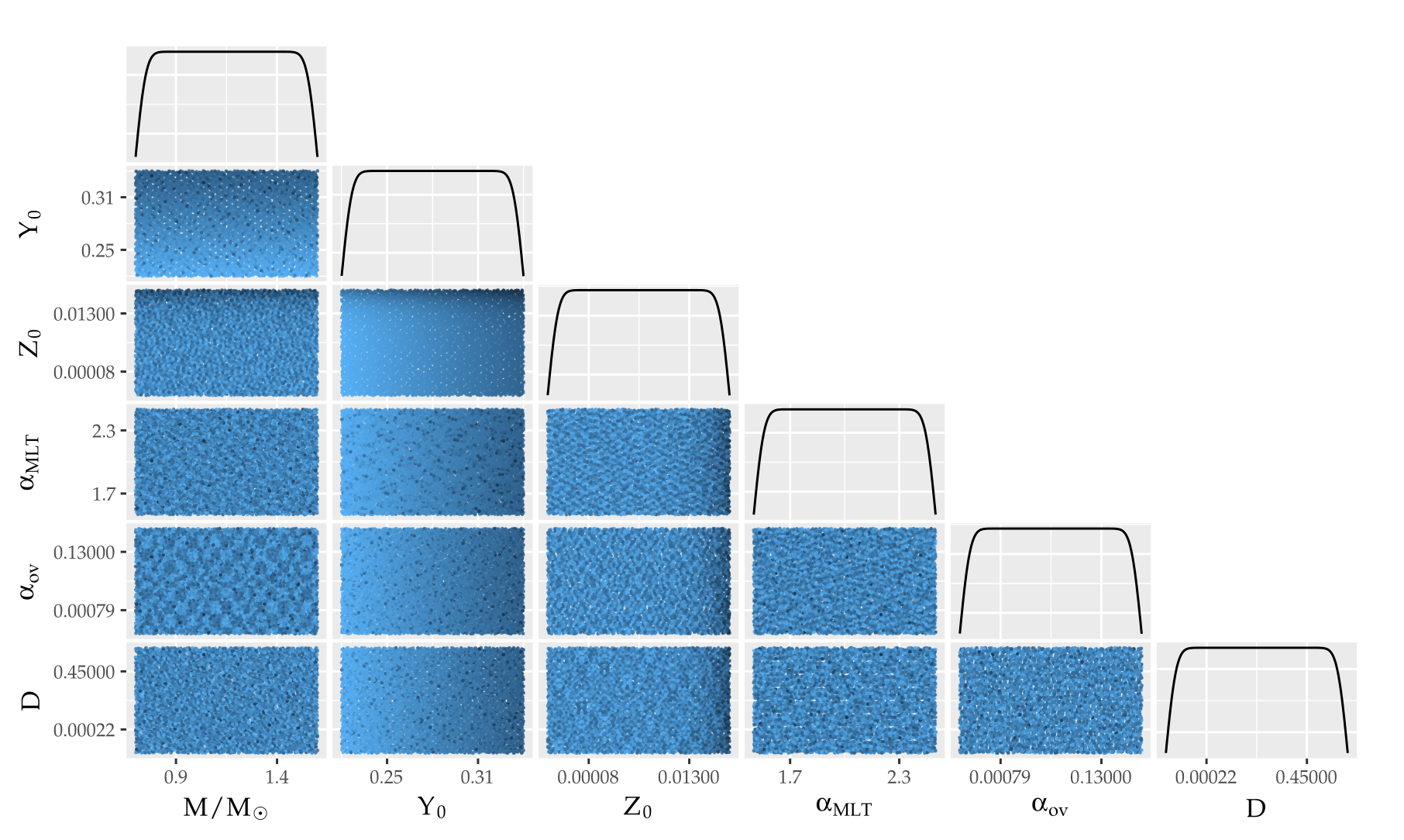}
    \caption{Scatterplot matrix (lower panels) and density plots (diagonal) of evolutionary track initial conditions considered. Mass (M), initial helium (Y$_0$), initial metallicity (Z$_0$), mixing length parameter ($\alpha_{\text{MLT}}$), overshoot ($\alpha_{\text{ov}}$), and diffusion multiplication factor (D) were varied in a quasi-random fashion to obtain a low-discrepancy grid of model tracks. Points are colored by their initial hydrogen X$_0=1-$Y$_0-$Z$_0$, with \mb{black} being low X$_0$ ($\approx 56\%$) and \mb{blue} being high X$_0$ ($\approx 78\%$). The parameter space is densely populated with evolutionary tracks of maximally different initial conditions. }
    \label{fig:inputs}
\end{figure*}

We use MESA version r8118 with the Helmholtz-formulated equation of state that allows for radiation pressure and interpolates within the 2005 update of the OPAL EOS tables \citep{2002ApJ...576.1064R}. We assume a \citet{1998SSRv...85..161G} solar composition for our initial abundances and opacity tables. Since we restrict our study to the main sequence, we use an eight-isotope nuclear network consisting of $^1$H, $^3$He, $^4$He, $^{12}$C, $^{14}$N, $^{16}$O, $^{20}$Ne, and $^{24}$Mg. We use a step function for overshooting and set a scaling factor $f_0 = \alpha_{\text{ov}}/5$ to determine the radius $r_0 = H_p \cdot f_0$ inside the convective zone at which convection switches to overshooting, where $H_p$ is the pressure scale height. \mb{The overshooting parameter applies to all convective boundaries and is kept fixed throughout the course of a track's evolution, so a non-zero value does not imply that the model has a convective core at any specific age.} 
All pre-main-sequence (PMS) models are calculated with a simple photospheric approximation, after which an Eddington T-$\tau$ atmosphere is appended on at ZAMS. We call ZAMS the point at which the nuclear luminosity of the models make up 99.9\% of the total luminosity. We calculate atomic diffusion with gravitation settling and without radiative levitation on the main sequence using five diffusion class representatives: $^1$H, $^3$He, $^4$He, $^{16}$O, and $^{56}$Fe \citep{burgers1969flow}.\footnote{The atomic number of each representative isotope is used to calculate the diffusion rate of the other isotopes allocated to that group; see \citet{Paxton2011}.} 
Following their most recent measurements, we correct the defaults in MESA of the gravitational constant ($G=6.67408\times 10^{-8}$ \si{\per\g\cm\cubed\per\square\s}; \citealt{2015arXiv150707956M}), the gravitational mass of the Sun (M$_\odot = 1.988475\times 10^{33}$ \si{\g} $= \mu G^{-1} = 1.32712440042\times 10^{11}$ \si{\km\per\s} $G^{-1}$, where $\mu$ is the standard gravitational parameter; \citealt{pitjeva2015determination}), and the solar radius (R$_\odot = 6.95568\times 10^{10}$ \si{\cm}; \citealt{2008ApJ...675L..53H}). 

Each track is evolved from ZAMS to either an age of $\tau=16$ Gyr or until terminal-age main sequence (TAMS), which we define as having a fractional core hydrogen abundance (X$_{\text{c}}$) below $10^{-3}$. Evolutionary tracks with efficient heavy-element settling can develop discontinuities in their surface abundances if they lack sufficient model resolution. We implement adaptive remeshing by recomputing any track with abundance discontinuities in its surface layers using finer spatial and temporal resolutions (see Appendix \mb{\ref{sec:remeshing}} for details). Running stellar physics codes in a batch mode like this requires care, so we manually inspect multiple evolutionary diagnostics to ensure that proper convergence has been achieved. 


\subsection{Calculation of Seismic Parameters}
\label{sec:seis}
\begin{figure*}
    \centering
    \includegraphics[width=0.5\linewidth,keepaspectratio]{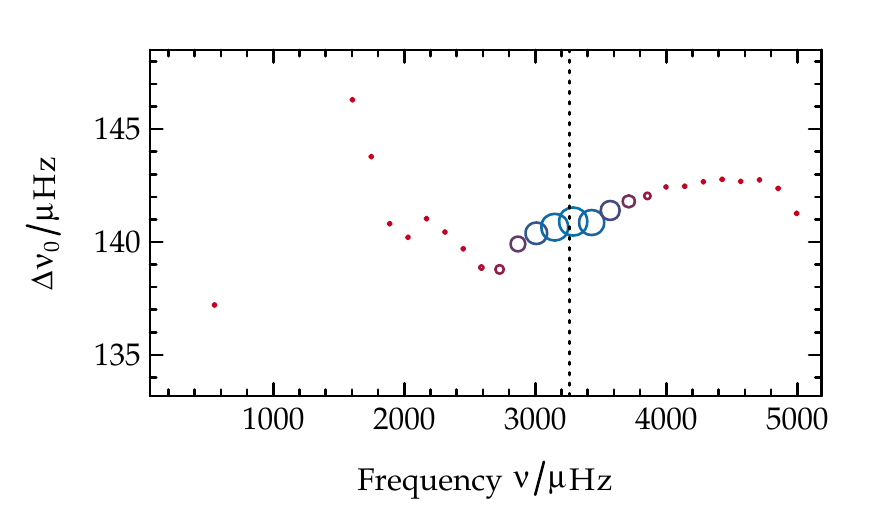}\hfill
    \includegraphics[width=0.5\linewidth,keepaspectratio]{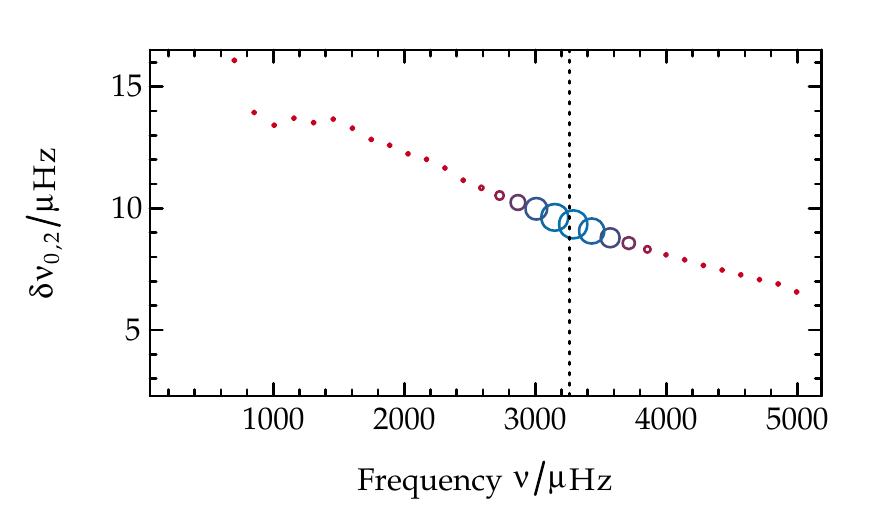}\\
    \includegraphics[width=0.5\linewidth,keepaspectratio]{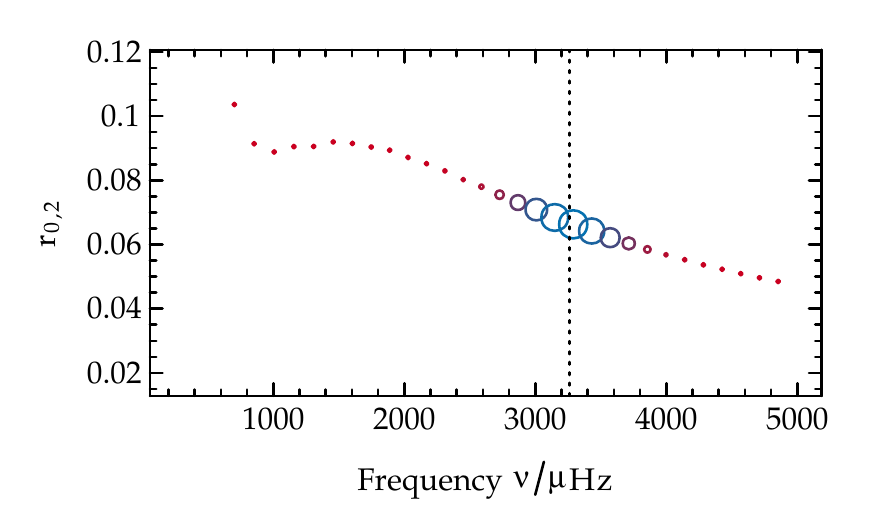}\hfill
    \includegraphics[width=0.5\linewidth,keepaspectratio]{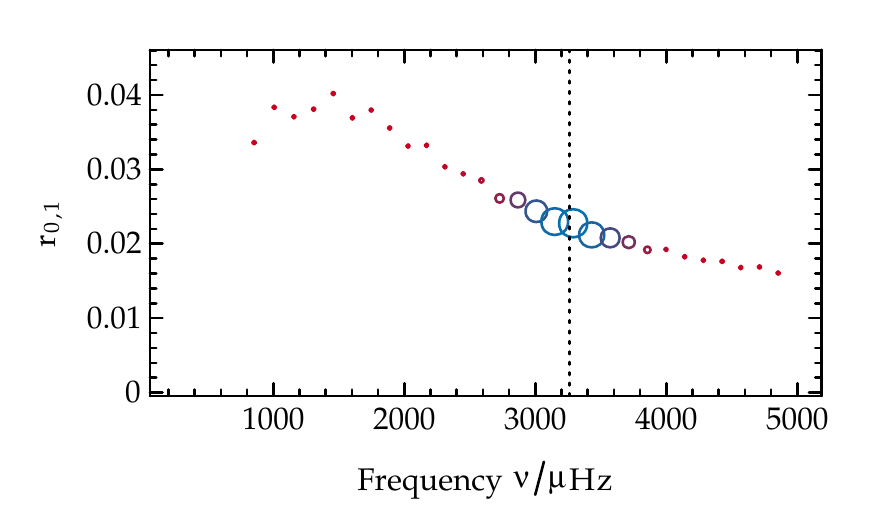}\\
    \caption{Calculation of seismic parameters for a stellar model. 
    The large and small frequency separations $\Delta\nu_0$ (top left) and $\delta\nu_{0,2}$ (top right) and frequency ratios $r_{0,2}$ (bottom left) and $r_{0,1}$ (bottom right) are shown as a function of frequency. The vertical dotted line in these bottom four plots indicates $\nu_{\max}$. Points are sized and colored proportionally to the applied weighting\mb{, with large blue symbols indicating high weight and small red symbols indicating low weight.} }%
    \label{fig:ratios}
\end{figure*}

We use the ADIPLS pulsation package \citep{2008ApSS.316..113C} to compute p-mode oscillations up to spherical degree $\ell=3$ below the acoustic cut-off frequency. We use on average of around 4\,000 points per stellar model and therefore have adequate resolution to calculate frequencies without remeshing. We denote any frequency separation $S$ as the difference between a frequency $\nu$ of spherical degree $\ell$ and radial order $n$ and another frequency, that is: 
\begin{equation} 
  S_{(\ell_1, \ell_2)}(n_1, n_2) \equiv \nu_{\ell_1}(n_1) - \nu_{\ell_2}(n_2).
\end{equation}
The large frequency separation is then
\begin{equation} 
  \Delta\nu_\ell(n) \equiv S_{(\ell, \ell)}(n, n-1)
\end{equation}
and the small frequency separation is
\begin{equation}
  \delta\nu_{(\ell, \ell+2)}(n) \equiv S_{(\ell, \ell+2)}(n, n-1).
\end{equation}
Near-surface layers of stars are poorly-modeled, which induces systematic frequency offsets \citep[see e.g.][]{1999A&A...351..689R}. The ratios between the large and small frequency separations (Equation \ref{eqn:LSratio}), and also between the large frequency separation and five-point-averaged frequencies (Equation \ref{eqn:rnl}) have been shown to be less sensitive to the surface term than the aforementioned separations and are therefore valuable asteroseismic diagnostics of stellar interiors \citep{2003A&A...411..215R}. They are defined as
\begin{equation} 
  \mathrm{r}_{(\ell,\ell+2)}(n) \equiv \frac{\delta\nu_{(\ell, \ell+2)}(n)}{\Delta\nu_{(1-\ell)}(n+\ell)} \label{eqn:LSratio}
\end{equation}
\begin{equation} 
  \mathrm{r}_{(\ell, 1-\ell)}(n) \equiv \frac{\mathrm{dd}_{(\ell,1-\ell)}(n)}{\Delta\nu_{(1-\ell)}(n+\ell)} \label{eqn:rnl}
\end{equation}
where
\begin{align} 
  \mathrm{dd}_{0,1}(n) \equiv \frac{1}{8} \big[&\nu_0(n-1) - 4\nu_1(n-1) 
                                 +6\nu_0(n) \notag\\&- 4\nu_1(n) + \nu_0(n+1)\big]\\ 
  \mathrm{dd}_{1,0}(n) \equiv -\frac{1}{8} \big[&\nu_1(n-1) - 4\nu_0(n) 
                                 +6\nu_1(n) \notag\\&- 4\nu_0(n+1) + \nu_1(n+1)\big].
\end{align}
Since the set of radial orders that are observable differs from star to star, we collect global statistics on $\Delta\nu_0$, $\delta\nu_{0,2}$, $\delta\nu_{1,3}$, $r_{0,2}$, $r_{1,3}$, $r_{0,1}$, and $r_{1,0}$. We mimic the range of observable frequencies in our models by weighting all frequencies by their position in a Gaussian envelope centered at the predicted frequency of maximum oscillation power $\nu_{\max}$ and having full-width at half-maximum of $0.66\cdot\nu_{\max}{}^{0.88}$ as per the prescription given by \citet{2012A&A...537A..30M}. We then calculate the weighted median of each variable, which we denote with angled parentheses (e.g.\ $\langle r_{0,2}\rangle$). We choose the median rather than the mean because it is a robust statistic with a high breakdown point, meaning that it is much less sensitive to the presence of outliers (for a discussion of breakdown points, see \citealt{hampel1971general}, who attributed them to Gauss). This approach allows us to predict the fundamental \mb{stellar} parameters of any solar-like oscillator with multiple observed modes irrespective of which exact radial orders have been detected. Illustrations of the methods used to derive the frequency separations and ratios of a stellar model are shown in Figure \ref{fig:ratios}. 

\subsection{Training the Random Forest} \label{sec:forest}
\begin{figure*}[ht]
    \centering
    \includegraphics[trim={0.9cm 0 0 0}]{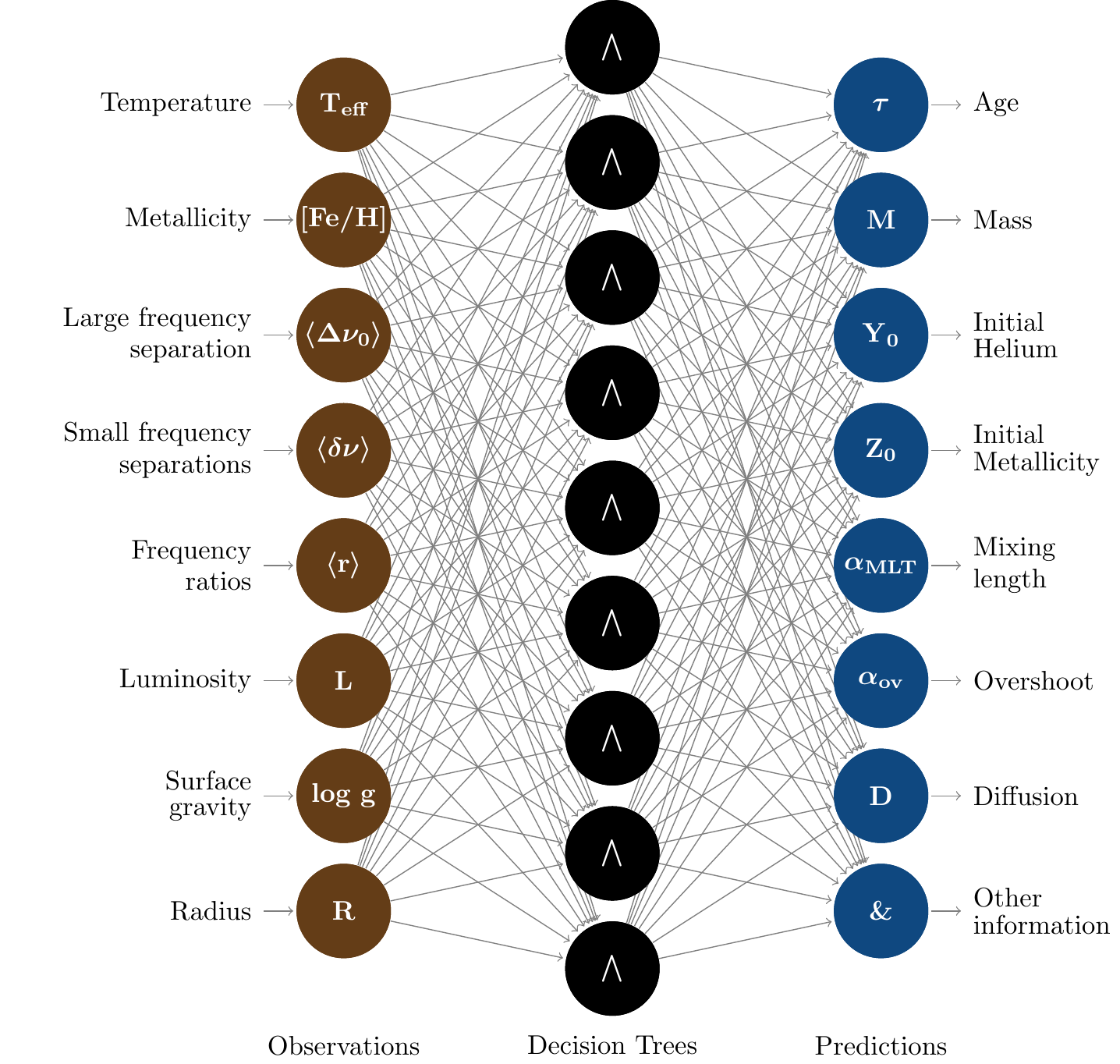}
    \caption{A schematic representation of a random forest regressor for inferring fundamental stellar parameters. \mb{Observable quantities} such as \mb{T$_{\text{eff}}$ and [Fe/H]} and global asteroseismic observables like \mb{$\langle\Delta\nu\rangle$ and} $\langle\delta\nu_{0,2}\rangle$ are input on the left side. These quantities are then fed through to some number of hidden decision trees, which each independently predict \mb{parameters} like age and mass. The predictions are then averaged and output on the right side. All inputs and outputs are optional. For example, surface gravities, luminosities, and radii are not always available \mb{from observations} (e.g.\ with the KOI stars\mb{, see Section \ref{sec:koi} below}). In their absence, these quantities can be predicted instead of being supplied. In this case, those nodes can be moved over to the ``prediction'' side instead of being on the ``observations'' side. Also, in addition to potentially unobserved inputs like stellar radii, other interesting model parameters can be predicted as well, such as core hydrogen mass fraction or surface helium abundance. \label{fig:rf} }
\end{figure*}

We train a random forest regressor on our matrix of evolutionary models to discover the relations that facilitate inference of stellar parameters from \mb{observed} quantities. A schematic representation of the topology of our random forest regressor can be seen in Figure \ref{fig:rf}. \mb{Random forests arise in machine learning through the family of algorithms known as CART, i.e. Classification and Regression Trees.} There are several good textbooks that discuss random forests \citep[see e.g.][Chapter 15]{hastie2005elements}. \mb{A random forest is an ensemble regressor, meaning that it is composed of many individual components that each perform statistical regression, and the forest subsequently averages over the results from each component \citep{breiman2001random}. The components of the ensemble are decision trees, each of which learns a set of decision rules for relating \mb{observable quantities} to \mb{stellar parameters}. An ensemble approach is preferred because using only a single decision tree that is able to see all of the training data may result in a regressor that has memorized the training data and is therefore unable to generalize to as yet unseen values. This undesirable phenomenon is known in machine learning as over-fitting, and is analogous to fitting $n$ data points using a degree $n$ polynomial: the fit will work perfectly on the data that was used for fitting, but fail badly on any unseen data. To avoid this, each decision tree in the forest is given a random subset of the evolutionary models and a random subset of the observable quantities from which to build a set of rules relating observed quantities to stellar parameters. This process, known as statistical bagging \citep[][Section 8.7]{hastie2005elements}, prevents the collection of trees from becoming over-fit to the training data, and thus results in a regression model that is capable of generalizing the information it has learned and predicting values for data on which it has not been trained. } 

\subsubsection{Feature Importance} \label{sec:importances}

\mb{The CART algorithm uses} information theory to decide which rule is the best choice for inferring \mb{stellar parameters} like age and mass from the supplied information \citep[][Chapter 9]{hastie2005elements}. At every stage, the rule that creates the largest decrease in mean squared error (MSE) is crafted. A rule may be, for example, ``all models with L $<0.4$ L$_\odot$ have M $<$ 1 M$_\odot$.'' Rules are created until every \mb{stellar model} that was supplied to that particular tree is fully explained by a sequence of decisions. We moreover use a variant on random forests known as \emph{extremely} randomized trees \citep{geurts2006extremely}, which further randomize attribute splittings (e.g.\ split on L) and the location of the cut-point (e.g.\ split on 0.4 L/L$_\odot$) used when creating decision rules. 

\begin{figure}
    \centering
    \includegraphics[width=\colwidth, keepaspectratio]{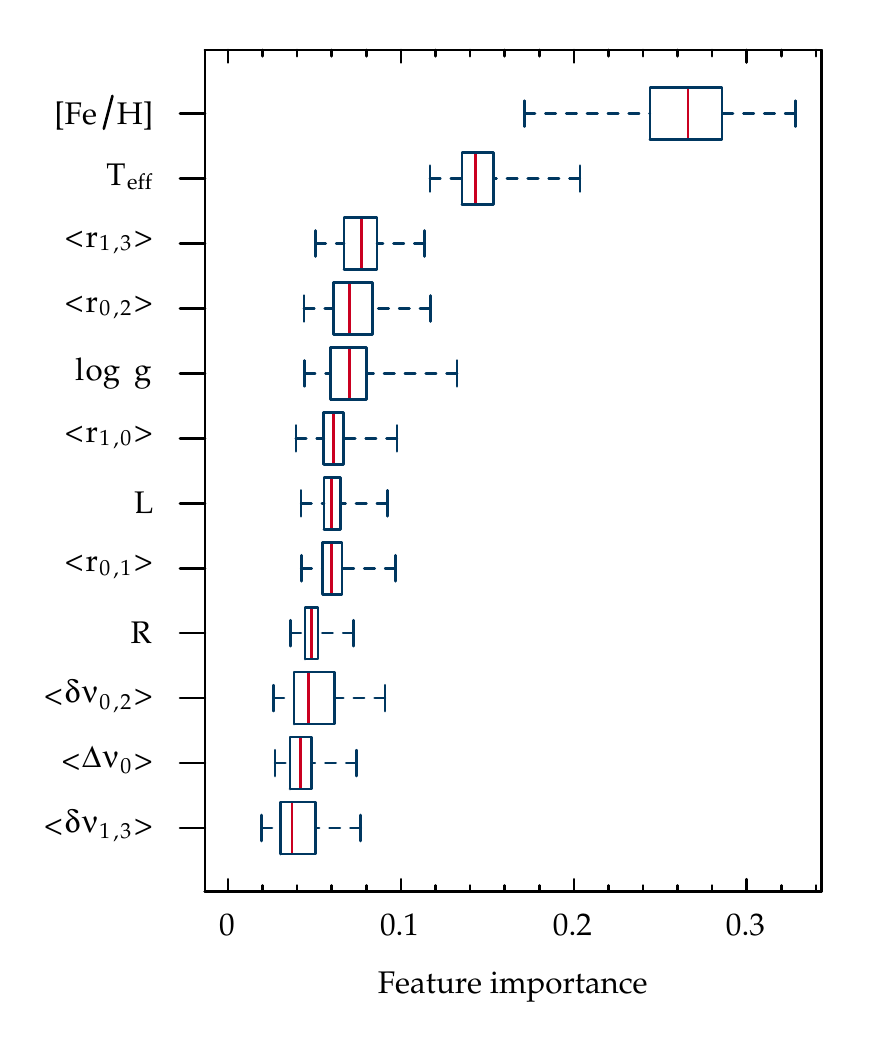}
    \caption{Relative importance of each observable feature in inferring fundamental stellar parameters as measured by a random forest regressor grown from a grid of evolutionary models. The boxes display the first (16\%) and third (84\%) quartile of feature importance over all trees, the center line indicates the median, and the whiskers extend to the most extreme values.}
    \label{fig:importances}
\end{figure}

\mb{The process of constructing a random forest} presents an opportunity for not only inferring stellar parameters from observations, but also for understanding the relationships that exist in the \mb{stellar models}. Each decision tree explicitly ranks the relative ``importance'' of each observable quantity \mb{for inferring stellar parameters}, where importance is defined in terms of both the reduction in MSE after defining a decision rule based on that quantity and the number of models that use that rule. \mb{In machine learning, the variables that have been measured and are supplied as inputs to the algorithm are known as ``features.'' Figure \ref{fig:importances} shows a feature importance plot, i.e.~distributions of relative importance over all of the trees in the forest for each feature used to infer stellar parameters. The features that are used most often to construct decision rules are metallicity and temperature, which are each significantly more important features than the rest.} The importance of [Fe/H] is due to the fact that the determinations of quantities like the Z$_0$ and D depend nearly entirely on it \citep[see also][]{main-sequence-stats}. Note that importance does not indicate indispensability: an appreciable fraction of decision rules being made based off of \mb{one feature} does not mean that another forest without that \mb{feature} would not perform just as well. That being said, these results indicate that the best area to improve measurements would be in metallicity determinations, because for stars being predicted using this random forest, less precise values here means exploring many more paths and hence arriving at less certain predictions.

\begin{figure*}[!hbtp]
    \centering
    \includegraphics[width=0.5\linewidth,keepaspectratio]{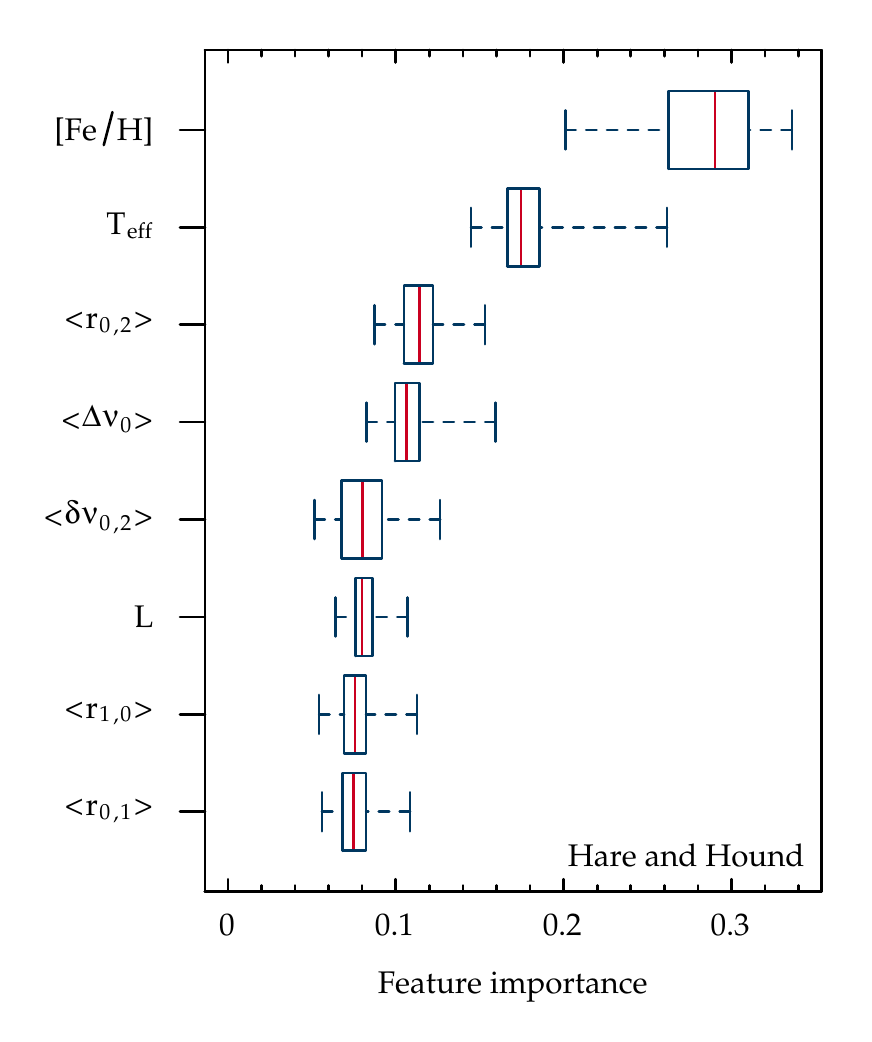}\hfill
    \includegraphics[width=0.5\linewidth, keepaspectratio]{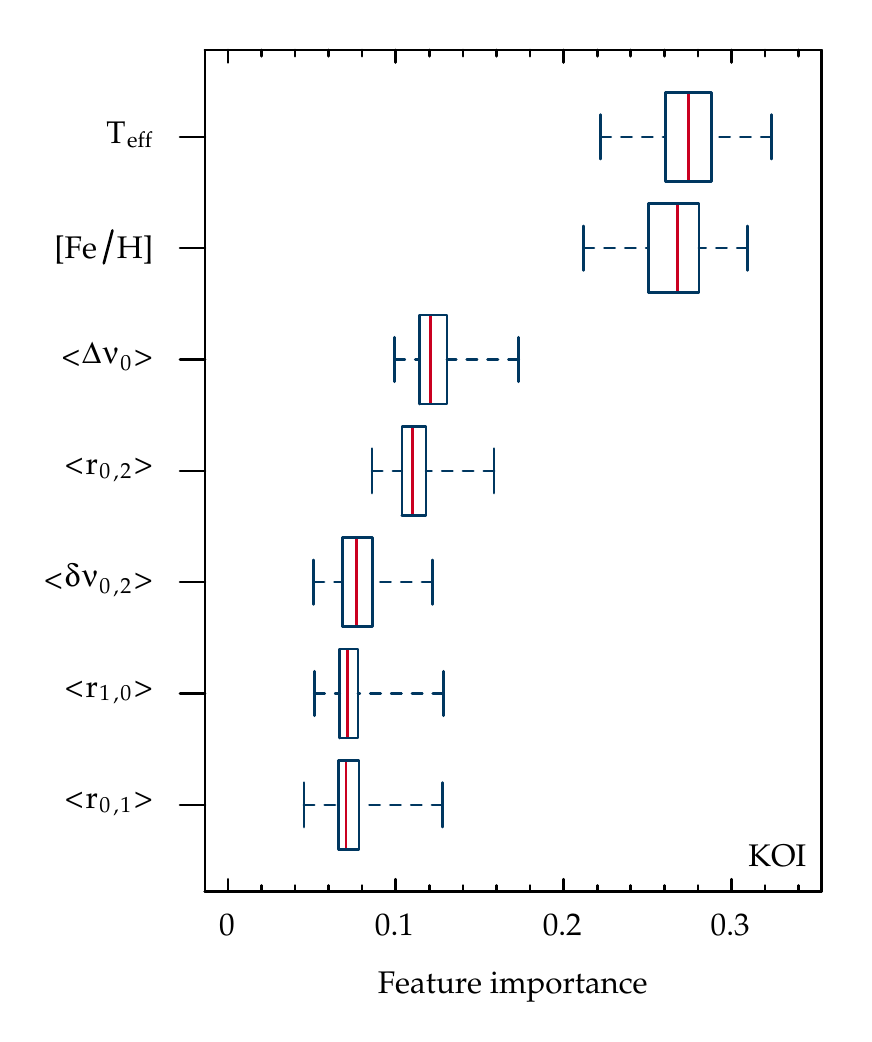}
    \caption{Box-and-whisker plots of relative importance for each observable feature in measuring fundamental stellar parameters for the hare-and-hound exercise data (left), where luminosities are available; and the \emph{Kepler} objects-of-interest (right), where they are not. Octupole ($\ell=3$) modes have not been measured in any of these stars, so $\langle\delta\nu_{1,3}\rangle$ and $\langle r_{1,3}\rangle$ from evolutionary modelling are not supplied to these random forests. The boxes are sorted by median importance.
    \label{fig:importances2} }
\end{figure*}

For many stars, \mb{stellar quantities} such as radii, luminosities, surface gravities, and/or oscillation modes with spherical degree $\ell=3$ are not available from observations. For example, the KOI data set \mb{(see Section \ref{sec:koi} below)} lacks all of this information, and the hare-and-hound exercise data \mb{(see Section \ref{sec:hnh} below)} lack all of these except luminosities. We therefore must train random forests that predict those quantities instead of using them as \mb{features}. We show the relative importance for \mb{the remaining features that were} used to train these forests in Figure \ref{fig:importances2}. When $\ell=3$ modes and luminosities are omitted, effective temperature jumps in importance and ties with [Fe/H] as the most \mb{important feature}. 

\subsubsection{Advantages of CART}
We choose random forests over any of the many other non-linear regression routines (e.g.\ Gaussian processes, symbolic regression, neural networks, support vector regression, etc.) for several reasons. First, random forests perform \emph{constrained} regression; that is, they only make predictions within the boundaries of the supplied training data \citep[see e.g.][Section 9.2.1]{hastie2005elements}. This is in contrast to other methods like neural networks, which ordinarily perform unconstrained regression and are therefore not prevented from predicting non-physical quantities such as negative masses or from violating conservation requirements. 

Secondly, due to the decision rule process that is explained below, random forests are insensitive to the scale of the data. Unless care is taken, other regression methods will artificially weight some observable \mb{quantities} like temperature as being more important than, say, luminosity, solely because temperatures are written using larger numbers (e.g.\ 5777 vs.\ 1, see for example section 11.5.3 of \citealt{hastie2005elements} for a discussion). Consequently, solutions obtained by other methods will change if they are \mb{run using features that are} expressed using different units of measure. For example, other methods will produce different regressors if trained on luminosity values expressed in solar units verses values expressed in erg\mb{s}, whereas random forests will not. \mb{Commonly, this problem is mitigated in other methods by means of variable standardization and through the use of Mahalabonis distances \citep{mahalanobis1936generalized}. However, these transformations are arbitrary, and handling variables naturally without rescaling is thus preferred. } 

Thirdly, random forests take only seconds to train, which can be a large benefit if different stars have different \mb{features} available. For example, some stars have luminosity information available whereas others do not, so a different regressor must be trained for each. In the extreme case, if one wanted to make predictions for stars using all of their respectively observed frequencies, one would need to train a new regressor for each star using the subset of simulated frequencies that correspond to the ones observed for that star. Ignoring the difficulties of surface-term corrections and mode identifications, such an approach would be well-handled by random forest, suffering only a small hit to performance from its relatively small training cost. On the other hand, it would be infeasible to do this on a star-by-star basis with most other routines such as deep neural networks, because \mb{those} methods can take days or even weeks to train. 

And finally\mb{, as we saw in the previous section,} random forests provide the opportunity to extract insight about the actual regression being performed by examining the importance of each \mb{feature} in making predictions.

\subsubsection{Uncertainty}
\label{sec:uncertainties}
There are three separate sources of uncertainty in predicting stellar parameters. The first is the systematic uncertainty in the physics used to model stars. These uncertainties are unknown, however, and hence cannot be propagated. The second is the uncertainty belonging to the observations of the star. We propagate measurement uncertainties $\sigma$ into the predictions by perturbing all measured quantities $n=10\,000$ times with normal noise having zero mean and standard deviation $\sigma$. We account for the covariance between asteroseismic separations and ratios by recalculating them upon each perturbation. 

The final source is regression uncertainty. Fundamentally, each parameter can only be constrained to the extent that observations are able to bear information pertaining to that parameter. Even if observations were error-free, there still may exist a limit to which information gleaned from the surface may tell us about the physical \mb{qualities} and evolutionary history of a star. We quantify those limits via cross-validation: we train the random forest on only a subset of the simulated evolutionary tracks and make predictions on a held-out validation set. We randomly hold out a different subset of the tracks 25 times to serve as different validation sets and obtain averaged accuracy scores.

We calculate accuracies using several scores. The first is the explained variance score V$_{\text{e}}$:
\begin{equation}
  \text{V}_{\text{e}} = 1 - \frac{\text{Var}\{ y - \hat y \}}{\text{Var}\{ y \}}
\end{equation}
where $y$ is the \mb{true} value we want to predict from the validation set (e.g.\ stellar mass), $\hat y$ is the predicted value from the random forest, and Var is the variance, i.e.\ the square of the standard deviation. This score tells us the extent to which the regressor has reduced the variance in the parameter it is predicting. The value ranges from negative infinity, which would be obtained by a pathologically bad predictor; to one for a perfect predictor, which occurs if all of the values are predicted with zero error. 

The next score we consider is the residuals of each prediction, i.e.\ the absolute difference between the true value $y$ and the predicted value $\hat y$. Naturally, we want this value to be as low as possible. We also consider the precision of the regression $\hat \sigma$ by taking the standard deviation of predictions across all of the decision trees in the forest. Finally, we consider these scores together by calculating the distance of the residuals in units of precision, i.e.\ $\abs{\hat y - y} / \hat{\sigma}$. 

\begin{figure*}
    \centering
    \includegraphics[width=0.66\linewidth,keepaspectratio]{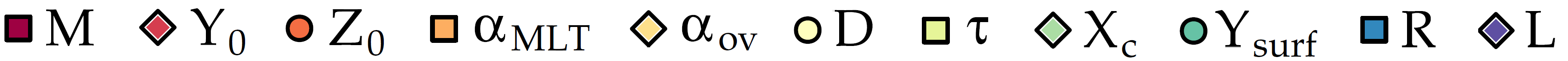}\\
    \includegraphics[width=0.5\linewidth,keepaspectratio]{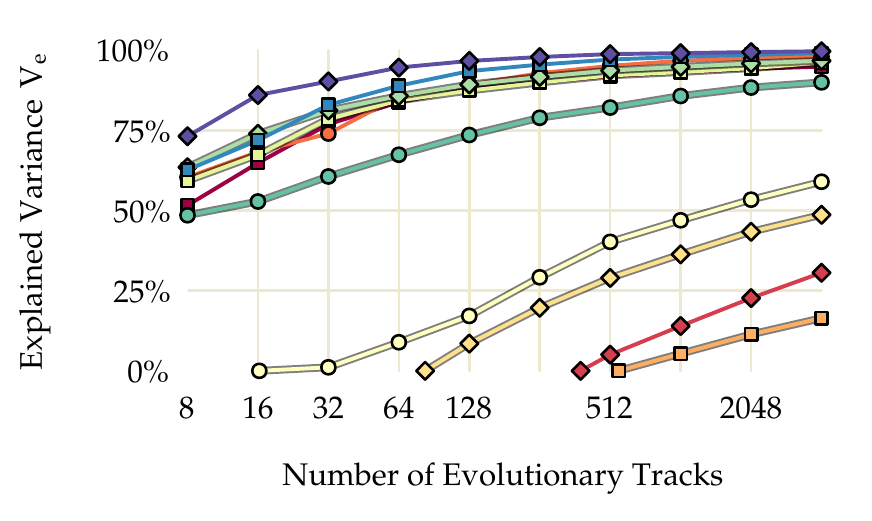}\hfill
    \includegraphics[width=0.5\linewidth,keepaspectratio]{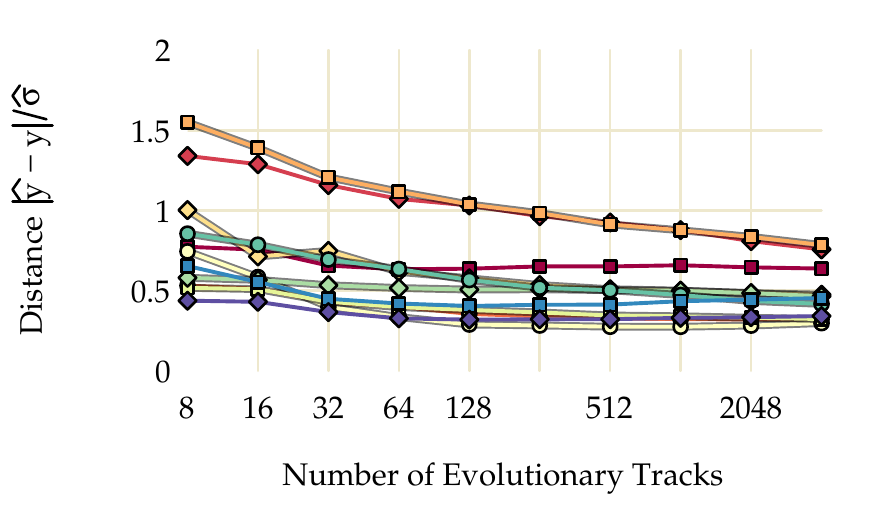}\\
    \includegraphics[width=0.5\linewidth,keepaspectratio]{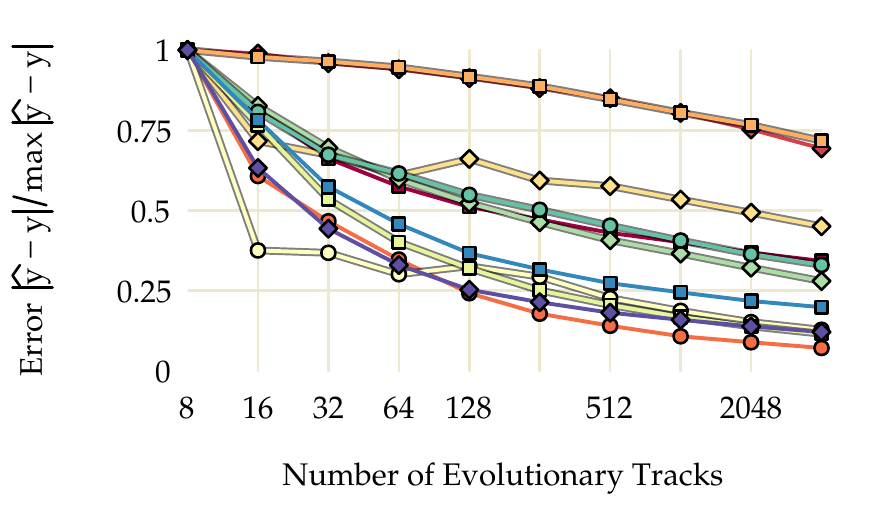}\hfill
    \includegraphics[width=0.5\linewidth,keepaspectratio]{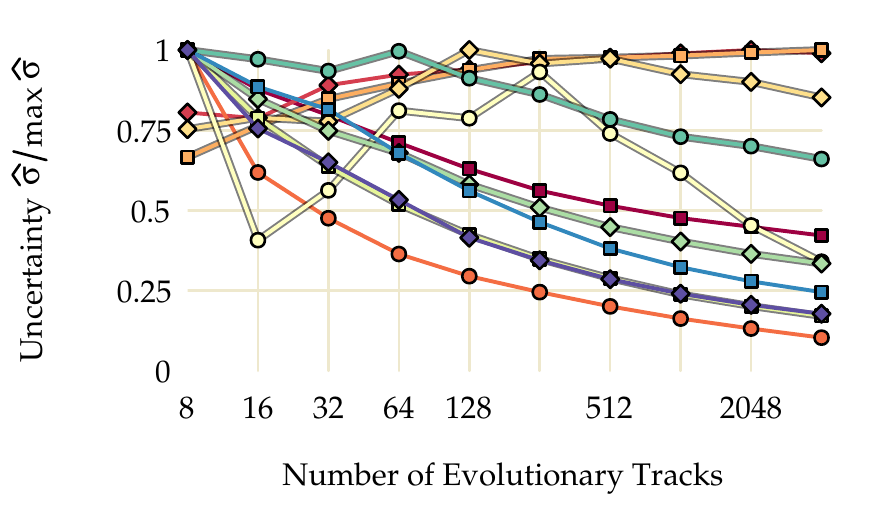}\\
    \caption{Evaluations of regression accuracy. Explained variance (top left), accuracy per precision distance (top right), normalized absolute error (bottom left), and normalized uncertainty (bottom right) for each stellar parameter as a function of the number of evolutionary tracks used in training the random forest. These results use 64 models per track and 256 trees in the random forest. } 
    \label{fig:evaluation-tracks}
\end{figure*}

Figure \ref{fig:evaluation-tracks} shows these accuracies as a function of the number of evolutionary tracks used in the training of the random forest. Since the residuals and standard deviations of each parameter are incomparable, we normalize them by dividing by the maximum value. We also consider the number of trees in the forest and the number of models per evolutionary track\mb{. In this work, we use 256 trees in each forest, which we have selected via cross-validation by choosing a number of trees that is greater than the point at which we saw that the explained variance was no longer increasing greatly;} see Appendix \ref{sec:evaluation} for an extended discussion. 

When supplied with enough \mb{stellar models}, the random forest reduces the variance in each parameter and is able to make precise inferences. The forest has very high predictive power for most \mb{parameters}, and as a result, essentially all of the uncertainty when predicting quantities such as stellar radii and luminosities will stem from observational uncertainty. However, for some model \mb{parameters}---most notably the mixing length parameter---there is still a great deal of variance in the residuals. Prior to \mb{the point where the regressor has been trained on} about 500 evolutionary tracks, the differences between the true and predicted mixing lengths actually have a greater variance than just the true mixing lengths themselves. Likewise, the diffusion multiplication factor is difficult to constrain because a star can achieve the same present-day [Fe/H] by either having a large initial non-hydrogen abundance and a large diffusion \mb{multiplication} factor, or by having the same initial [Fe/H] as present [Fe/H] but with diffusion disabled. These difficult-to-constrain \mb{parameters} will therefore be predicted with substantial \mb{uncertainties} regardless of the precision of the observations. 


\section{Results}
We perform three tests of our method. We begin with a hare-and-hound simulation exercise to show that we can reliably recover parameters. We then move to the Sun and the solar-like stars 16 Cyg A \& B, which have been the subjects of many investigations; and we conclude by applying our method to 34 \emph{Kepler} objects-of-interest. In each case, we train our random forest regressor on the subset of data that is available for the stars being processed. In the case of the Sun and 16 Cygni, we know very accurately their radii, luminosities, and surface gravities. For other stars, we will predict this information instead of supplying it.

\subsection{Hare and Hound} 
\label{sec:hnh}
\begin{figure}
    \centering
    \includegraphics[width=\colwidth,keepaspectratio]{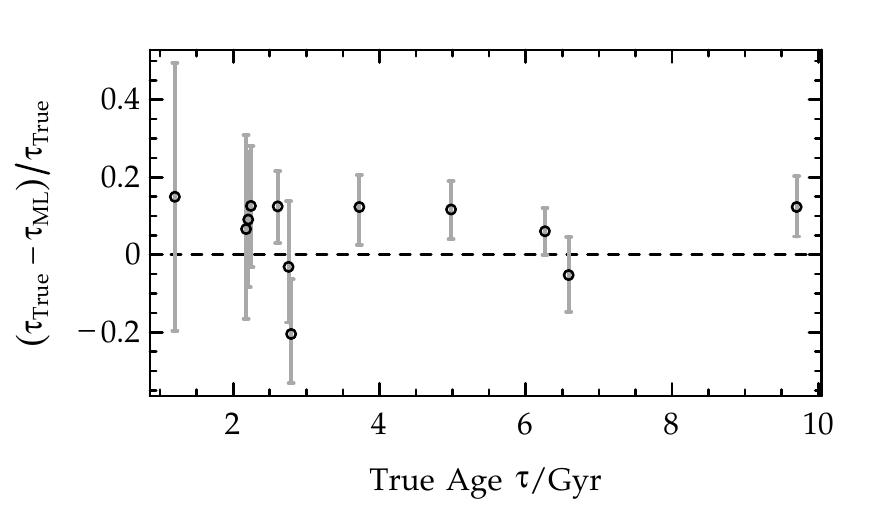}\\
    \includegraphics[width=\colwidth,keepaspectratio]{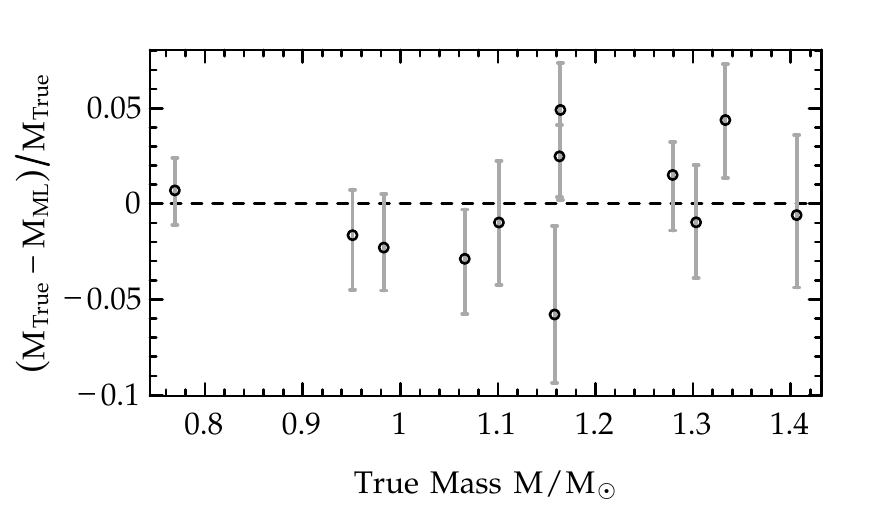}\\
    \includegraphics[width=\colwidth,keepaspectratio]{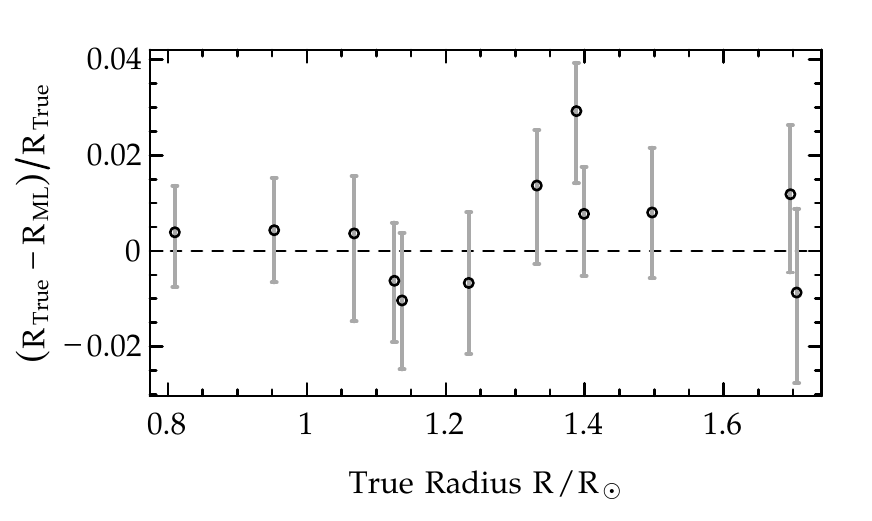}
    \caption{Relative differences between the predicted and true values for age (top), mass (middle), and radius (bottom) as a function of the true values in \mb{the} hare-and-hound simulation exercise. \vspace*{5mm} 
    \label{fig:hare-comparison}}
\end{figure}

We performed a blind hare-and-hound exercise to evaluate the performance of our predictor. Author S.B.\ prepared twelve models varied in mass, initial chemical composition, and mixing length parameter with only some models having overshooting and only some models having atomic diffusion included. The models were evolved without rotation using the Yale rotating stellar evolution code \citep[YREC;][]{2008ApSS.316...31D}, which is a different evolution code than the one that was used to train the random forest. Effective temperatures, luminosities, [Fe/H] and $\nu_{\max}$ values as well as $\ell=0,1,2$ frequencies were obtained from each model. Author G.C.A.\  perturbed the ``observations'' of these models according to the scheme devised by \citet{spaceinn}. \mb{Appendix \ref{sec:hare-and-hound} lists the true values and the perturbed observations of the hare-and-hound models}. The perturbed observations and their uncertainties were given to author E.P.B.\@, who used the described method to recover the stellar parameters of \mb{these} models without being given access to the true values. Relative differences between the true and predicted ages, masses, and radii for these models are plotted against their true values in Figure \ref{fig:hare-comparison}. The method is able to recover the true model values within uncertainties even when they have been perturbed by noise. We do not compare the predicted mixing length parameter, overshooting parameter or diffusion \mb{multiplication} factor \mb{the interpretation of these parameters depends on how they have been defined and their precise implementation.}

\subsection{The Sun and 16 Cygni}
\begin{figure*}
    \centering
    \includegraphics[width=\linewidth,keepaspectratio]{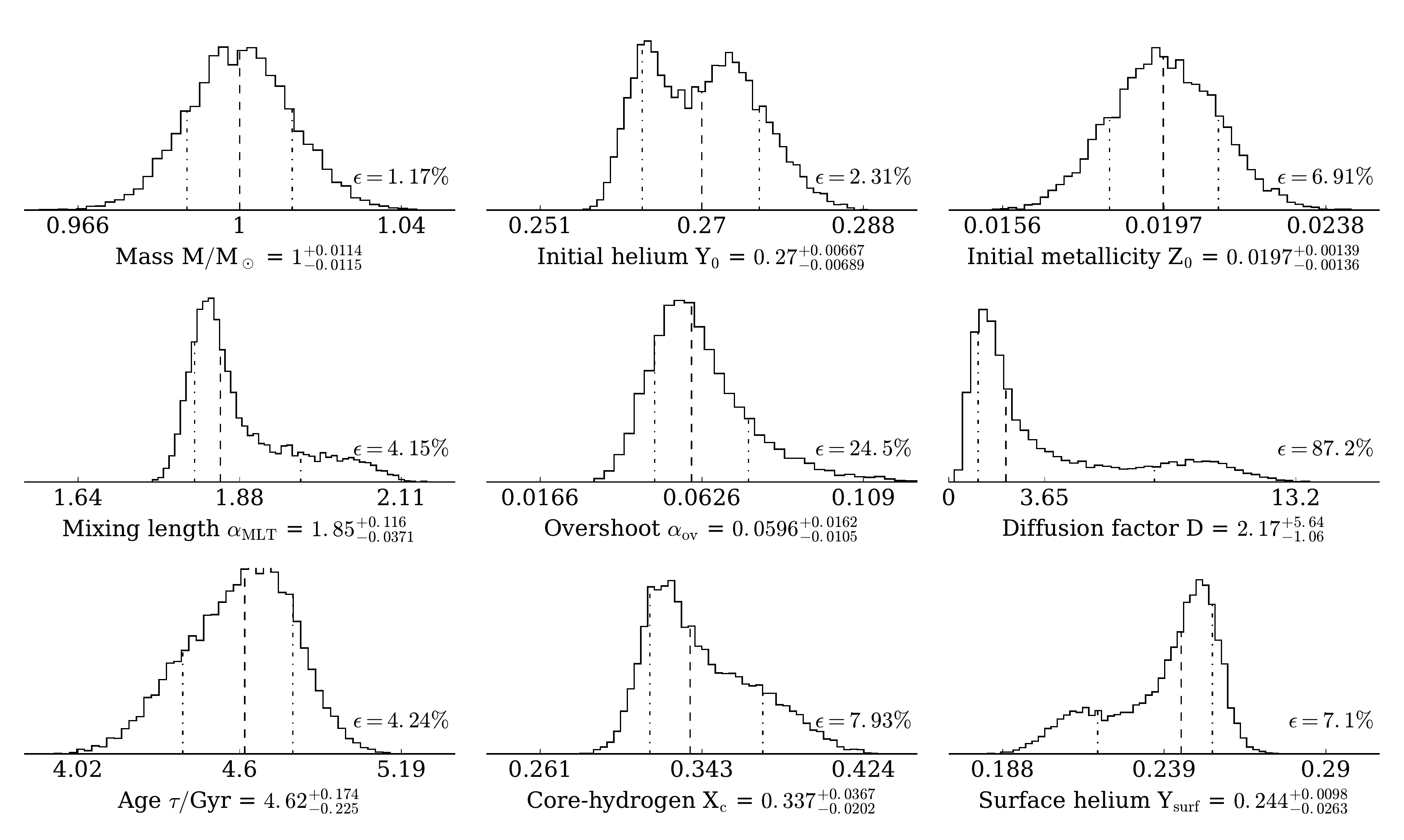}
    \caption{Predictions from machine learning of initial (top six) and current (bottom three) stellar parameters for degraded solar data. Labels are placed at the mean and 3$\sigma$ levels. \mb{Dashed and dot-dashed} lines indicate the median and quartiles\mb{, respectively}. Relative uncertainties $\epsilon$ are shown beside each plot. Note that the overshoot parameter applies to all convective boundaries and is not modified over the course of evolution, so a non-zero value does not imply a convective core. 
    \label{fig:corner} } 
\end{figure*}

To ensure confidence in our predictions on \emph{Kepler} data, we first degrade the frequencies of the Sun at solar minimum that were obtained by the Birmingham Solar-Oscillations Network \citep[BiSON;][]{2014MNRAS.439.2025D} to the level of information that is achievable by the spacecraft. We also degrade the Sun's uncertainties of \mb{other} observations by applying 16 Cyg B's uncertainties of effective temperature, luminosity, surface gravity, metallicity, $\nu_{\max}$, radius, and radial velocity. Finally, we perturb each value with random Gaussian noise according to its uncertainty to reflect the fact that the measured value of an uncertain observation is not \emph{per se} the true value. We use the random forest whose feature importances were shown in Figure \ref{fig:importances} to predict the values of the Sun; i.e.\ the random forest trained on effective temperatures, metallicities, luminosities, surface gravities, radii, and asteroseismic \mb{quantities} $\langle \Delta\nu_0 \rangle$, $\langle \delta\nu_{0,2} \rangle$, $\langle \delta\nu_{1,3} \rangle$, $\langle r_{0,2} \rangle$, $\langle r_{1,3} \rangle$, $\langle r_{0,1} \rangle$, and $\langle r_{1,0} \rangle$. We show in Figure \ref{fig:corner} the densities for the predicted mass, initial composition, mixing length parameter, overshoot coefficient, and diffusion multiplication factor needed for fitting an evolutionary model to degraded data of the Sun as well as the predicted solar age, core hydrogen abundance, and surface helium abundance. \mb{As discussed in Section \ref{sec:uncertainties}, these densities show the distributions resulting from running $10\,000$ different noise perturbations fed through the random forest.} Relative uncertainties $\epsilon=100\cdot\sigma/\mu$ are also indicated, where $\mu$ is the mean and $\sigma$ is the standard deviation of the quantity being predicted. Our predictions are in good agreement with the known values (see also Table \ref{tab:results} and Table \ref{tab:results-ca}). 

Several parameters show multimodality due to model degeneracies. For example, two solutions for the initial helium are present. This is because it covaries with the mixing length parameter: the peak of \mb{higher} Y$_0$ corresponds to the peak of \mb{lower} $\alpha_{\text{MLT}}$ and vice versa. Likewise, high values of surface helium correspond to low values of the diffusion \mb{multiplication} factor. 

Effective temperatures, surface gravities, and metallicities of 16 Cyg A and B were obtained from \citet{2009A&A...508L..17R}; radii and luminosities from \citet{2013MNRAS.433.1262W}; and frequencies from \citet{2015MNRAS.446.2959D}. We obtained the radial velocity measurements of 16 Cyg A and B from \citet{2002ApJS..141..503N} and corrected frequencies for Doppler shifting as per the prescription in \citet{2014MNRAS.445L..94D}. We tried with and without line-of-sight corrections and found that it did not affect the predicted quantities or their uncertainties. We use the same random forest as we used for the degraded solar data to predict the \mb{parameters} of these stars. The initial parameters---masses, chemical compositions, mixing lengths, diffusion \mb{multiplication} factors, and overshoot coefficients---for 16 Cygni as predicted by machine learning \mb{are shown} in Table \ref{tab:results}, and the predicted current parameters---age, surface helium and core hydrogen abundances---\mb{are shown} in Table \ref{tab:results-ca}. For reference we also show the predicted solar values from these inputs there as well. These results support the hypothesis that 16 Cyg A and B were co-natal; i.e.\ they formed at the same time with the same initial composition. 

\begin{figure*}
    \centering
    \includegraphics[width=0.5\linewidth, keepaspectratio]{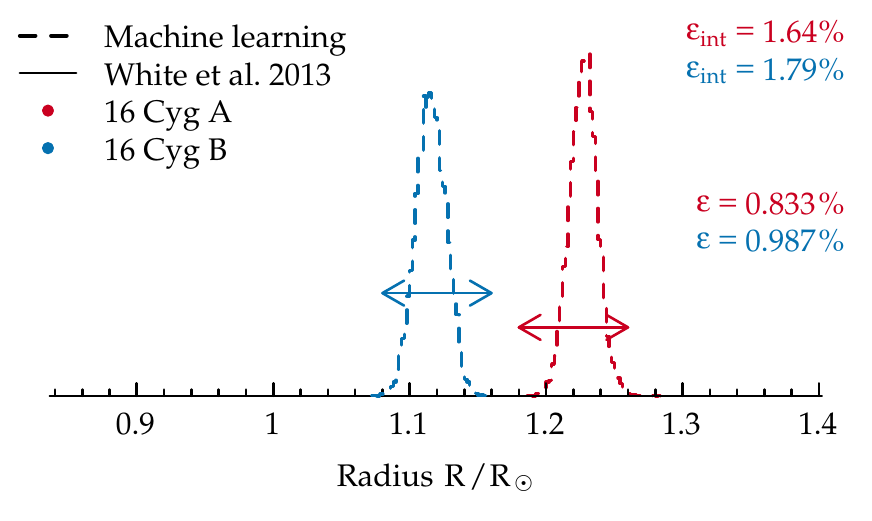}\hfill
    \includegraphics[width=0.5\linewidth, keepaspectratio]{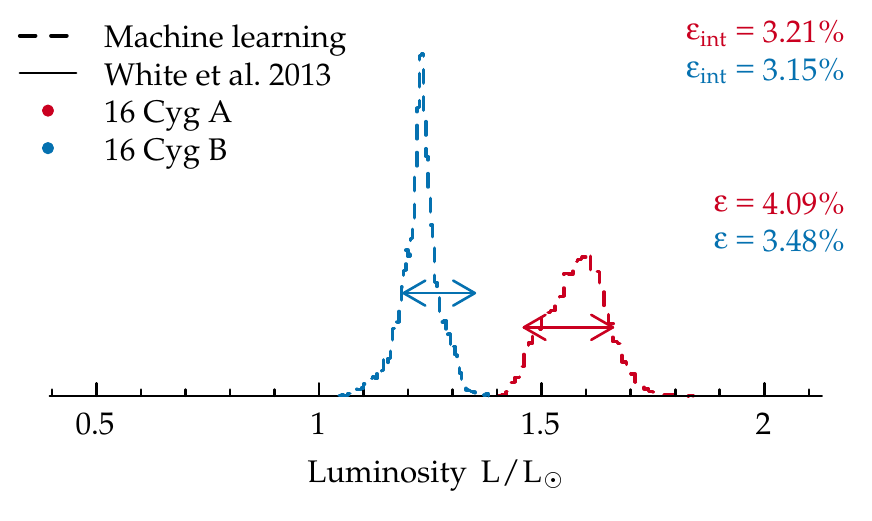}\\
    \includegraphics[width=0.5\linewidth, keepaspectratio]{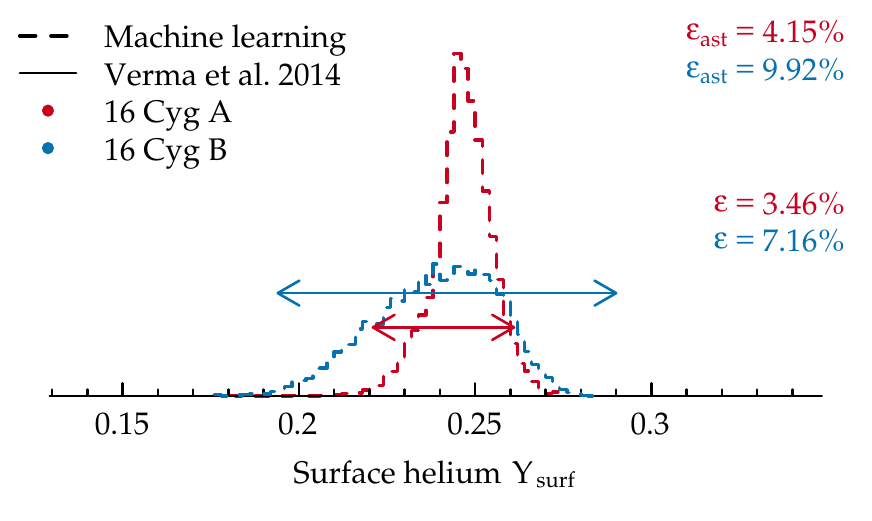}
    \caption{Probability densities for predictions of 16 Cyg A (red) and B (blue) from machine learning of radii (top left), luminosities (top right), and surface helium abundances (bottom). Relative uncertainties $\epsilon$ are shown beside each plot. Predictions and $2\sigma$ uncertainties from interferometric (``int'') measurements and asteroseismic (``ast'') estimates are shown with arrows.}
    \label{fig:interferometry}
\end{figure*}

We additionally predict the radii and luminosities of 16 Cyg A and B instead of using them as \mb{features}. Figure \ref{fig:interferometry} shows our inferred radii, luminosities and surface helium abundances of 16 Cyg A and B plotted \mb{along with} the values determined by interferometry \citep{2013MNRAS.433.1262W} and an asteroseismic estimate \citep{2014ApJ...790..138V}. Here again we find excellent agreement between our method and the measured values. 

\begin{figure*}
    \centering
    \includegraphics[width=0.5\linewidth, keepaspectratio]{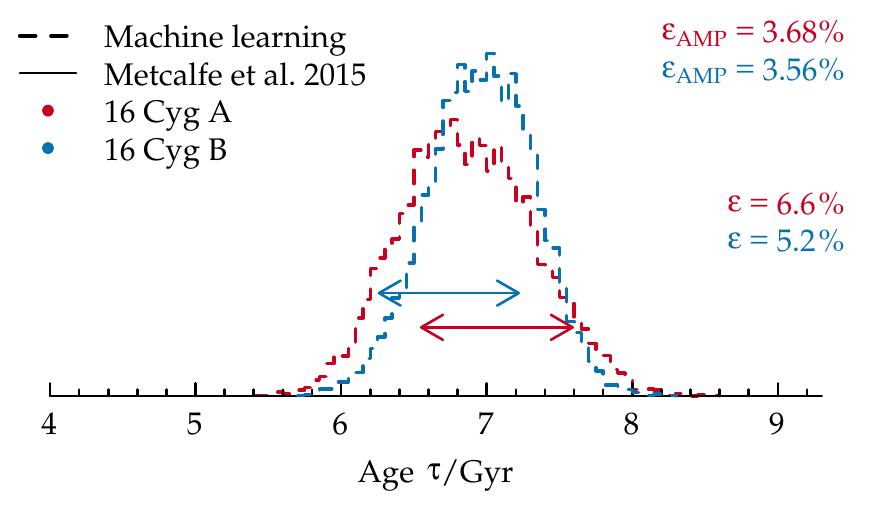}\hfill
    \includegraphics[width=0.5\linewidth, keepaspectratio]{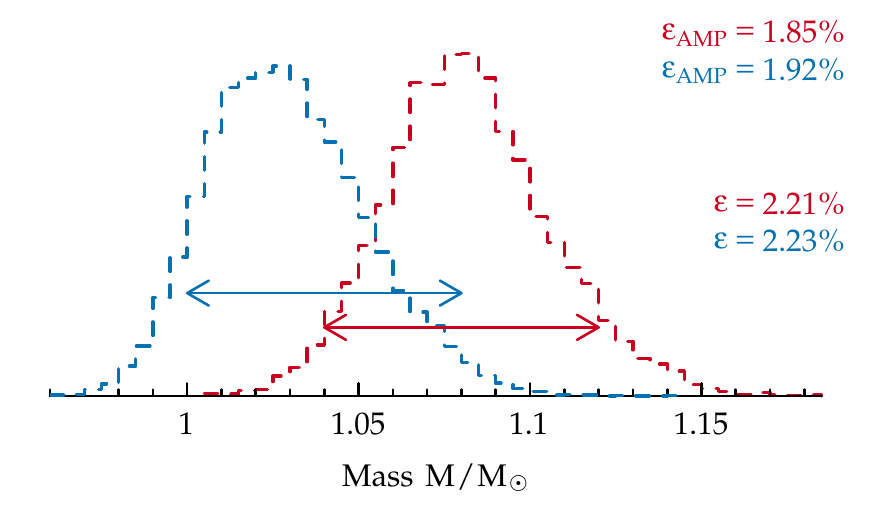}\\
    \includegraphics[width=0.5\linewidth, keepaspectratio]{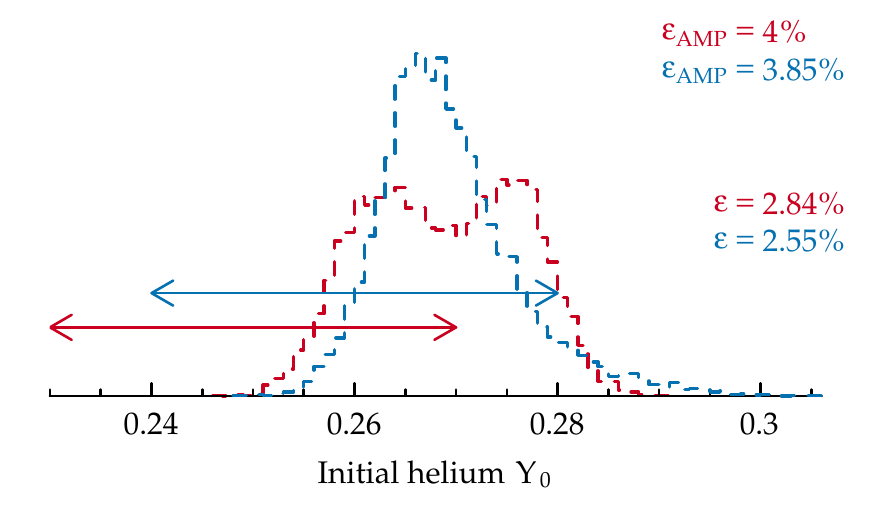}\hfill
    \includegraphics[width=0.5\linewidth, keepaspectratio]{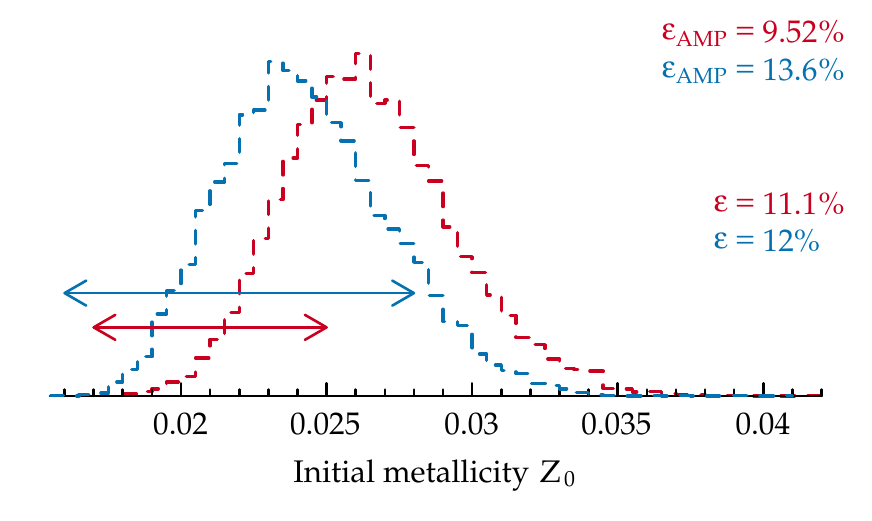}
    \caption{Probability densities showing predictions from machine learning of fundamental stellar parameters for 16 Cyg A (red) and B (blue) \mb{along with} predictions from AMP modelling. Relative uncertainties are shown beside each plot. Predictions and $2\sigma$ uncertainties from AMP modelling are shown with arrows. \vspace*{5mm}
    \label{fig:16Cyg-hist}}
\end{figure*}

\citet{2015ApJ...811L..37M} performed detailed modelling of 16 Cyg A and B using the Asteroseismic Modeling Portal (AMP), a genetic algorithm for matching individual frequencies of stars to stellar models. They calculated their results without heavy-element diffusion (i.e.\ with helium-only diffusion) and without overshooting. In order to account for systematic uncertainties, they multiplied the spectroscopic uncertainties of 16 Cyg A and B by an arbitrary constant $C=3$. Therefore, in order to make a fair comparison between the results of our method and theirs, we generate a new matrix of evolutionary models with those same conditions and also increase the uncertainties on [Fe/H] by a factor of $C$. In Figure \ref{fig:16Cyg-hist}, we show probability densities of the predicted parameters of 16 Cyg A and B that we obtain using machine learning in comparison with the results obtained by AMP. We find the values and uncertainties agree well. To perform their analysis, AMP required more than $15\,000$ hours of CPU time to model 16 Cyg A and B using the world's 10th fastest supercomputer, the Texas Advanced Computing Center Stampede \citep{TOP500}. Here we have obtained comparable results in roughly one minute \mb{on a computing cluster with 64 2.5 GHz cores} using only global asteroseismic \mb{quantities} and no individual frequencies. Although more computationally expensive than our method, detailed optimization codes like AMP do have advantages in that they are additionally able to obtain detailed structural models of stars. 

\begin{deluxetable*}{cccccccc}
\tablecaption{Means and standard deviations for predicted initial stellar parameters of the Sun (degraded data) and 16 Cyg A and B. 
\label{tab:results}}
\tablewidth{0pt}
\tablehead{\colhead{Name} & \colhead{M$/$M$_\odot$} & \colhead{Y$_0$} & \colhead{Z$_0$} & \colhead{$\alpha_{\mathrm{MLT}}$} & \colhead{$\alpha_{\mathrm{ov}}$} & \colhead{D}}
\startdata
Sun      & 1.00 $\pm$ 0.012 & 0.270 $\pm$ 0.0062 & 0.020 $\pm$ 0.0014 & 1.88 $\pm$ 0.078 & 0.06 $\pm$ 0.015 & 3.7 $\pm$ 3.18 \\
16 Cyg A & 1.08 $\pm$ 0.016 & 0.262 $\pm$ 0.0073 & 0.022 $\pm$ 0.0014 & 1.86 $\pm$ 0.077 & 0.07 $\pm$ 0.028 & 0.9 $\pm$ 0.76 \\
16 Cyg B & 1.03 $\pm$ 0.015 & 0.268 $\pm$ 0.0065 & 0.021 $\pm$ 0.0015 & 1.83 $\pm$ 0.069 & 0.11 $\pm$ 0.029 & 1.9 $\pm$ 1.57
\enddata
\end{deluxetable*}

\begin{deluxetable*}{cccc}
\tablecaption{Means and standard deviations for predicted current-age stellar \mb{parameters} of the Sun (degraded data) and 16 Cyg A and B. \label{tab:results-ca}}
\tablewidth{0pt}
\tablehead{\colhead{Name} & \colhead{$\tau/$Gyr} & \colhead{X$_{\mathrm{c}}$} & \colhead{Y$_{\mathrm{surf}}$}}
\startdata 
Sun      & 4.6 $\pm$ 0.20 & 0.34 $\pm$ 0.027 & 0.24  $\pm$ 0.017 \\
16 Cyg A & 6.9 $\pm$ 0.40 & 0.06 $\pm$ 0.024 & 0.246 $\pm$ 0.0085 \\
16 Cyg B & 6.8 $\pm$ 0.28 & 0.15 $\pm$ 0.023 & 0.24  $\pm$ 0.017
\enddata
\end{deluxetable*}

\subsection{\emph{Kepler} Objects of Interest}
\label{sec:koi}
\begin{figure*}
    \centering
    \includegraphics[width=0.332\linewidth,keepaspectratio]{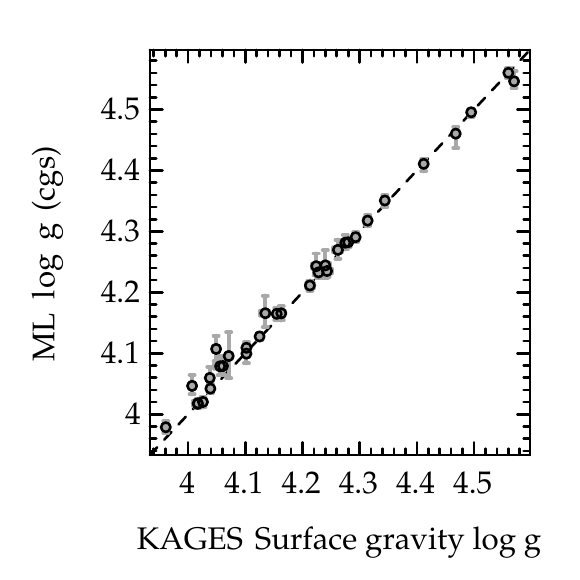}\hfill
    \includegraphics[width=0.332\linewidth,keepaspectratio]{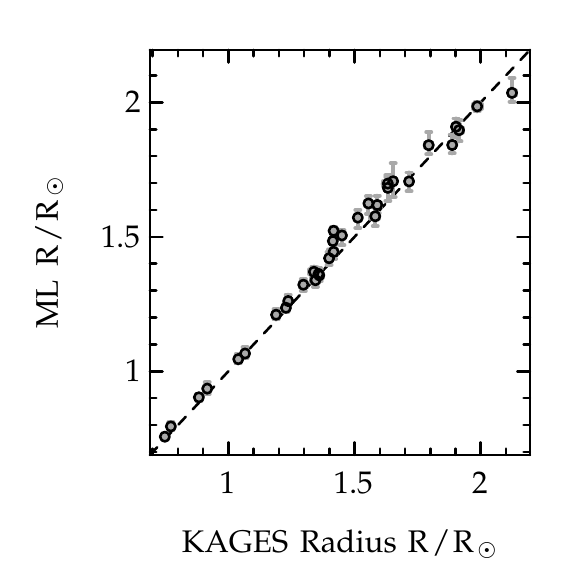}\hfill
    \includegraphics[width=0.332\linewidth,keepaspectratio]{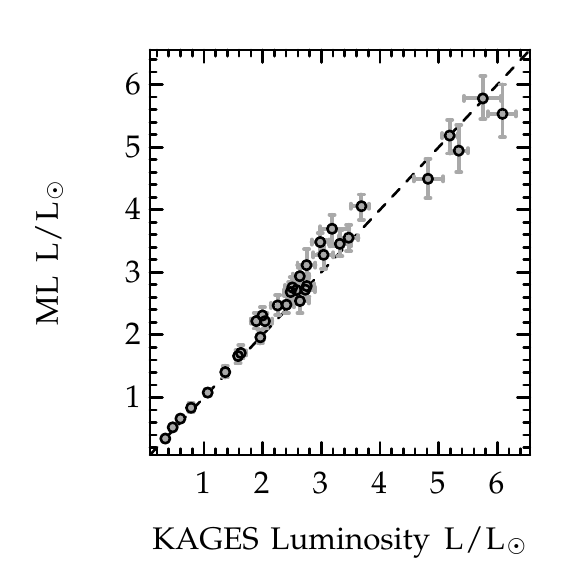}\\
    \includegraphics[width=0.332\linewidth,keepaspectratio]{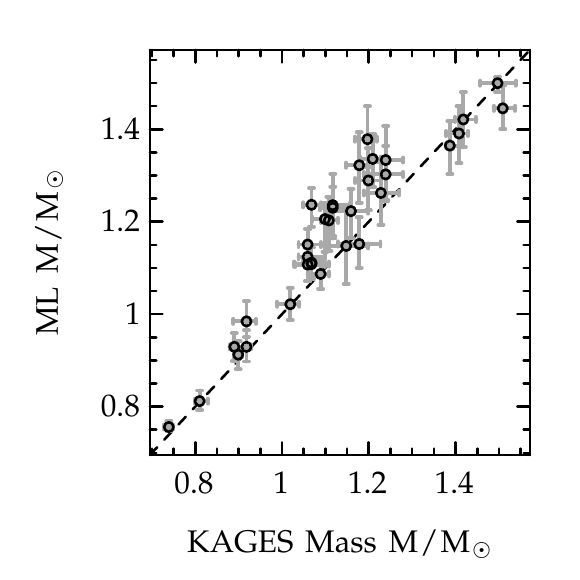}
    \includegraphics[width=0.332\linewidth,keepaspectratio]{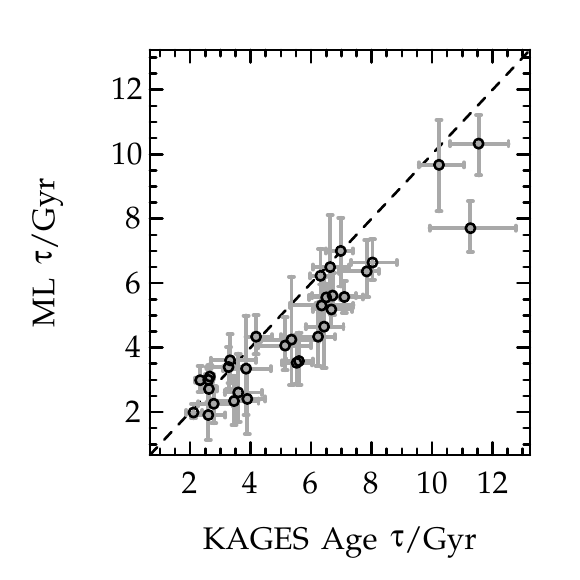}
    \caption{Predicted surface gravities, radii, luminosities, masses, and ages of 34 \emph{Kepler} objects-of-interest plotted against the suggested KAGES values. Medians, 16\% quantiles, and 84\% quantiles are shown for each point. A dashed line of agreement is shown in all panels to guide the eye. }
    \label{fig:us-vs-them}
\end{figure*}

We obtain observations and frequencies of the KOI targets from \citet{2016MNRAS.456.2183D}. We use line-of-sight radial velocity corrections when available, which was only the case for KIC 6278762 \citep{2002AJ....124.1144L}, KIC 10666592 \citep{2013A&A...554A..84M}, and KIC 3632418 \citep{2006AstL...32..759G}. We use the random forest whose feature importances were shown in Figure \ref{fig:importances2} to predict the fundamental \mb{parameters} of these stars; that is, the random forest that is trained on effective temperatures, metallicities, and asteroseismic quantities $\langle \Delta\nu_0 \rangle$, $\langle \delta\nu_{0,2} \rangle$, $\langle r_{0,2} \rangle$, $\langle r_{0,1} \rangle$, and $\langle r_{1,0} \rangle$. The predicted initial conditions---masses, chemical compositions, mixing lengths, overshoot coefficients, and diffusion \mb{multiplication} factors---are shown in Table \ref{tab:results-kages}; and the predicted current conditions---ages, core hydrogen abundances, surface gravities, luminosities, radii, and surface helium abundances---are shown in Table \ref{tab:results-kages-curr}. Figure \ref{fig:us-vs-them} shows the fundamental parameters obtained from our method plotted against those obtained by \citet[hereinafter KAGES]{2015MNRAS.452.2127S}. We find good agreement across all stars. 

Although still in statistical agreement, the median values of our predicted ages are systematically lower and the median values of our predicted masses are systematically higher than those predicted by KAGES. We conjecture that these discrepancies arise from differences in input physics. We vary the efficiency of diffusion, the extent of convective overshooting, and the value of the mixing length parameter to arrive at these estimates, whereas \mb{the KAGES} models are calculated using fixed \mb{amounts} of diffusion, without overshoot, and with a solar-calibrated mixing length. Models with overshooting, for example, will be more evolved at the same age due to having larger core masses. Without direct access to their models, however, the exact reason is difficult to pinpoint. 

\begin{figure*}
    \centering
    \includegraphics[width=\linewidth,keepaspectratio]{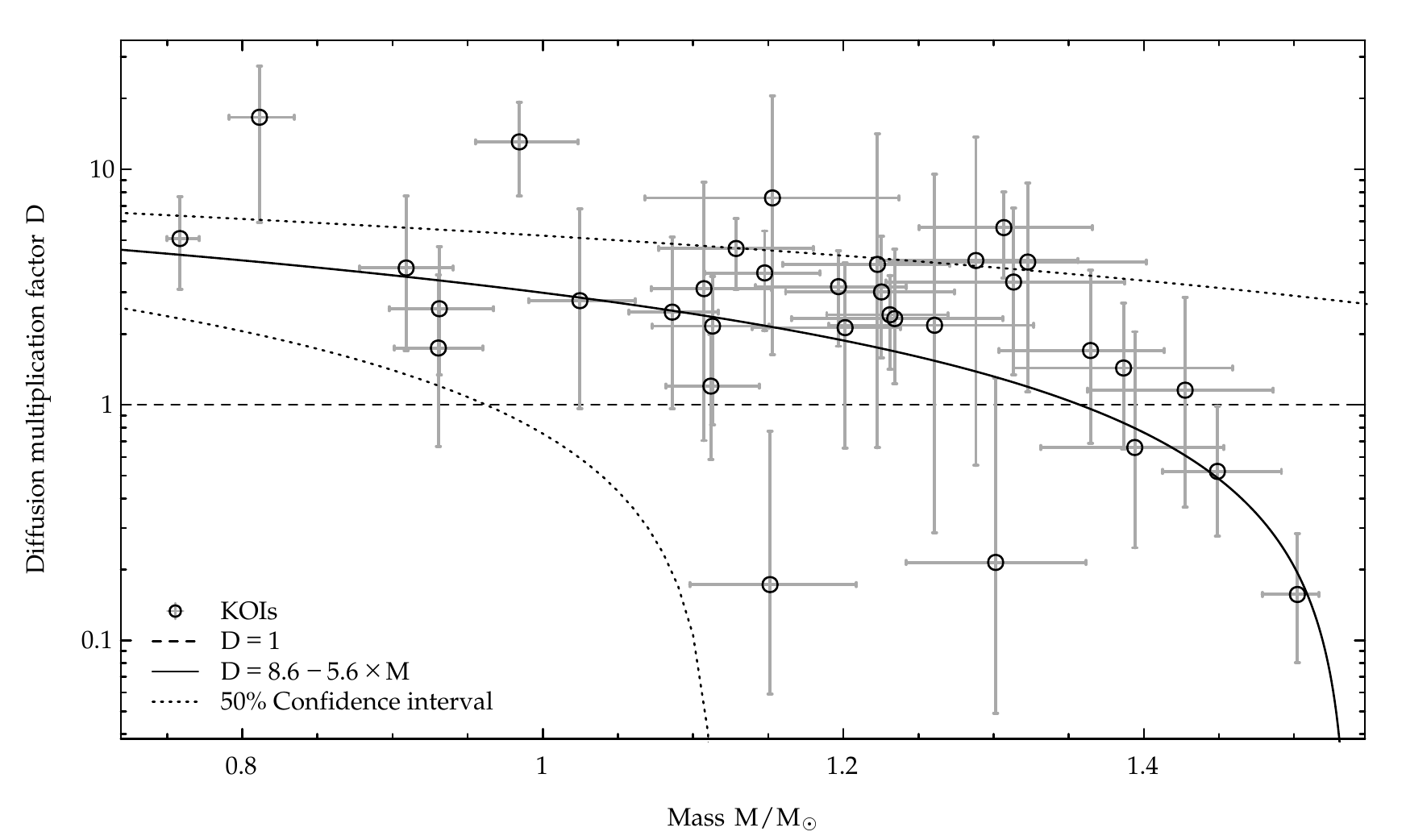}
    \caption{Logarithmic diffusion multiplication factor as a function of stellar mass for 34 \emph{Kepler} objects-of-interest. The solid line is the line of best fit from Equation \ref{eq:diffusion} and the dashed lines are the 50\% confidence interval around this fit. \label{fig:diffusion} } 
\end{figure*}

We find a significant linear trend in the \emph{Kepler} objects-of-interest between the diffusion multiplication factor and stellar mass needed to reproduce observations ($P = 0.0001$ from a two-sided t-test with $N-2=32$ degrees of freedom). Since the values of mass and diffusion \mb{multiplication} factor are uncertain, we use Deming regression to estimate the coefficients of this relation without regression dilution \citep{deming1943statistical}. We show the diffusion multiplication factors as a function of stellar mass for all of these stars in Figure \ref{fig:diffusion}. We find that the diffusion \mb{multiplication} factor linearly decreases with mass, i.e.\ 
\begin{equation} \label{eq:diffusion}
    \text{D} = ( 8.6 \pm 1.94 ) - ( 5.6 \pm 1.37 ) \cdot \text{M}/\text{M}_\odot
\end{equation}
and that this relation explains observations better than any constant factor (e.g.\ D=1 or D=0). 

\begin{deluxetable*}{ccccccc}
\tablecaption{Means and standard deviations for initial conditions of the KOI data set inferred via machine learning. The values obtained from degraded solar data predicted on these quantities are shown for reference. \label{tab:results-kages}}
\tablewidth{0pt}
\tablehead{\colhead{KIC} & \colhead{M$/$M$_\odot$} & \colhead{Y$_0$} & \colhead{Z$_0$} & \colhead{$\alpha_{\mathrm{MLT}}$} & \colhead{$\alpha_{\mathrm{ov}}$} & \colhead{D}}
\startdata
 3425851 & 1.15 $\pm$ 0.053  & 0.28  $\pm$ 0.020  & 0.015 $\pm$ 0.0028 & 1.9  $\pm$ 0.23  & 0.06 $\pm$ 0.057 &  0.5 $\pm$  0.92 \\
 3544595 & 0.91 $\pm$ 0.032  & 0.270 $\pm$ 0.0090 & 0.015 $\pm$ 0.0028 & 1.9  $\pm$ 0.10  & 0.2  $\pm$ 0.11  &  4.9 $\pm$  4.38 \\
 3632418 & 1.39 $\pm$ 0.057  & 0.267 $\pm$ 0.0089 & 0.019 $\pm$ 0.0032 & 2.0  $\pm$ 0.12  & 0.2  $\pm$ 0.14  &  1.1 $\pm$  1.01 \\
 4141376 & 1.03 $\pm$ 0.036  & 0.267 $\pm$ 0.0097 & 0.012 $\pm$ 0.0025 & 1.9  $\pm$ 0.12  & 0.1  $\pm$ 0.11  &  4.0 $\pm$  4.09 \\
 4143755 & 0.99 $\pm$ 0.037  & 0.277 $\pm$ 0.0050 & 0.014 $\pm$ 0.0026 & 1.77 $\pm$ 0.033 & 0.37 $\pm$ 0.071 & 13.4 $\pm$  5.37 \\
 4349452 & 1.22 $\pm$ 0.056  & 0.28  $\pm$ 0.012  & 0.020 $\pm$ 0.0043 & 1.9  $\pm$ 0.17  & 0.10 $\pm$ 0.090 &  7.3 $\pm$  8.82 \\
 4914423 & 1.19 $\pm$ 0.048  & 0.274 $\pm$ 0.0097 & 0.026 $\pm$ 0.0046 & 1.8  $\pm$ 0.11  & 0.08 $\pm$ 0.043 &  2.3 $\pm$  1.6 \\
 5094751 & 1.11 $\pm$ 0.038  & 0.274 $\pm$ 0.0082 & 0.018 $\pm$ 0.0030 & 1.8  $\pm$ 0.11  & 0.07 $\pm$ 0.041 &  2.3 $\pm$  1.39 \\
 5866724 & 1.29 $\pm$ 0.065  & 0.28  $\pm$ 0.011  & 0.027 $\pm$ 0.0058 & 1.8  $\pm$ 0.13  & 0.12 $\pm$ 0.086 &  7.0 $\pm$  8.38 \\
 6196457 & 1.31 $\pm$ 0.058  & 0.276 $\pm$ 0.005  & 0.032 $\pm$ 0.0050 & 1.71 $\pm$ 0.050 & 0.16 $\pm$ 0.055 &  5.7 $\pm$  2.34 \\
 6278762 & 0.76 $\pm$ 0.012  & 0.254 $\pm$ 0.0058 & 0.013 $\pm$ 0.0017 & 2.09 $\pm$ 0.069 & 0.06 $\pm$ 0.028 &  5.3 $\pm$  2.23 \\
 6521045 & 1.19 $\pm$ 0.046  & 0.273 $\pm$ 0.0071 & 0.027 $\pm$ 0.0044 & 1.82 $\pm$ 0.074 & 0.12 $\pm$ 0.036 &  3.2 $\pm$  1.31 \\
 7670943 & 1.30 $\pm$ 0.061  & 0.28  $\pm$ 0.017  & 0.021 $\pm$ 0.0045 & 2.0  $\pm$ 0.23  & 0.06 $\pm$ 0.064 &  1.0 $\pm$  2.55 \\
 8077137 & 1.23 $\pm$ 0.070  & 0.270 $\pm$ 0.0093 & 0.018 $\pm$ 0.0028 & 1.8  $\pm$ 0.14  & 0.2  $\pm$ 0.11  &  2.9 $\pm$  2.08 \\
 8292840 & 1.15 $\pm$ 0.079  & 0.28  $\pm$ 0.010  & 0.016 $\pm$ 0.0049 & 1.8  $\pm$ 0.15  & 0.1  $\pm$ 0.12  & 11.  $\pm$ 10.7  \\
 8349582 & 1.23 $\pm$ 0.040  & 0.271 $\pm$ 0.0069 & 0.043 $\pm$ 0.0074 & 1.9  $\pm$ 0.12  & 0.11 $\pm$ 0.060 &  2.5 $\pm$  1.11 \\
 8478994 & 0.81 $\pm$ 0.022  & 0.272 $\pm$ 0.0082 & 0.010 $\pm$ 0.0012 & 1.91 $\pm$ 0.054 & 0.21 $\pm$ 0.068 & 17.  $\pm$  9.74 \\
 8494142 & 1.42 $\pm$ 0.058  & 0.27  $\pm$ 0.010  & 0.028 $\pm$ 0.0046 & 1.70 $\pm$ 0.064 & 0.10 $\pm$ 0.051 &  1.6 $\pm$  1.65 \\
 8554498 & 1.39 $\pm$ 0.067  & 0.272 $\pm$ 0.0082 & 0.031 $\pm$ 0.0032 & 1.70 $\pm$ 0.077 & 0.14 $\pm$ 0.079 &  1.7 $\pm$  1.17 \\
 8684730 & 1.44 $\pm$ 0.030  & 0.277 $\pm$ 0.0075 & 0.041 $\pm$ 0.0049 & 1.9  $\pm$ 0.14  & 0.29 $\pm$ 0.094 & 15.2 $\pm$  8.81 \\
 8866102 & 1.26 $\pm$ 0.069  & 0.28  $\pm$ 0.013  & 0.021 $\pm$ 0.0048 & 1.8  $\pm$ 0.15  & 0.08 $\pm$ 0.070 &  5.  $\pm$  7.48 \\
 9414417 & 1.36 $\pm$ 0.054  & 0.264 $\pm$ 0.0073 & 0.018 $\pm$ 0.0028 & 1.9  $\pm$ 0.13  & 0.2  $\pm$ 0.1   &  2.2 $\pm$  1.68 \\
 9592705 & 1.45 $\pm$ 0.038  & 0.27  $\pm$ 0.010  & 0.029 $\pm$ 0.0038 & 1.72 $\pm$ 0.064 & 0.12 $\pm$ 0.056 &  0.6 $\pm$  0.47 \\
 9955598 & 0.93 $\pm$ 0.028  & 0.27  $\pm$ 0.011  & 0.023 $\pm$ 0.0039 & 1.9  $\pm$ 0.10  & 0.2  $\pm$ 0.13  &  2.2 $\pm$  1.76 \\
10514430 & 1.13 $\pm$ 0.053  & 0.277 $\pm$ 0.0046 & 0.021 $\pm$ 0.0039 & 1.78 $\pm$ 0.059 & 0.30 $\pm$ 0.097 &  4.7 $\pm$  1.77 \\
10586004 & 1.31 $\pm$ 0.078  & 0.274 $\pm$ 0.0055 & 0.038 $\pm$ 0.0071 & 1.8  $\pm$ 0.13  & 0.2  $\pm$ 0.13  &  4.3 $\pm$  3.99 \\
10666592 & 1.50 $\pm$ 0.023  & 0.30  $\pm$ 0.013  & 0.030 $\pm$ 0.0032 & 1.8  $\pm$ 0.11  & 0.06 $\pm$ 0.043 &  0.2 $\pm$  0.14 \\
10963065 & 1.09 $\pm$ 0.031  & 0.264 $\pm$ 0.0083 & 0.014 $\pm$ 0.0025 & 1.8  $\pm$ 0.11  & 0.05 $\pm$ 0.027 &  3.1 $\pm$  2.68 \\
11133306 & 1.11 $\pm$ 0.044  & 0.272 $\pm$ 0.0099 & 0.021 $\pm$ 0.0040 & 1.8  $\pm$ 0.16  & 0.04 $\pm$ 0.033 &  5.  $\pm$  5.75 \\
11295426 & 1.11 $\pm$ 0.033  & 0.27  $\pm$ 0.010  & 0.025 $\pm$ 0.0036 & 1.81 $\pm$ 0.084 & 0.05 $\pm$ 0.035 &  1.3 $\pm$  0.87 \\
11401755 & 1.15 $\pm$ 0.039  & 0.271 $\pm$ 0.0057 & 0.015 $\pm$ 0.0023 & 1.88 $\pm$ 0.055 & 0.33 $\pm$ 0.071 &  3.8 $\pm$  1.81 \\
11807274 & 1.32 $\pm$ 0.079  & 0.276 $\pm$ 0.0097 & 0.024 $\pm$ 0.0051 & 1.77 $\pm$ 0.083 & 0.11 $\pm$ 0.066 &  5.4 $\pm$  5.61 \\
11853905 & 1.22 $\pm$ 0.055  & 0.272 $\pm$ 0.0072 & 0.029 $\pm$ 0.0050 & 1.8  $\pm$ 0.12  & 0.18 $\pm$ 0.086 &  3.3 $\pm$  1.85 \\
11904151 & 0.93 $\pm$ 0.033  & 0.265 $\pm$ 0.0091 & 0.016 $\pm$ 0.0030 & 1.8  $\pm$ 0.13  & 0.05 $\pm$ 0.029 &  3.1 $\pm$  2.09 \\
     Sun & 1.00 $\pm$ 0.0093 & 0.266 $\pm$ 0.0035 & 0.018 $\pm$ 0.0011 & 1.81 $\pm$ 0.032 & 0.07 $\pm$ 0.021 &  2.1 $\pm$  0.83
\enddata
\end{deluxetable*}

\begin{deluxetable*}{ccccccc}
\tablecaption{Means and standard deviations for current-age conditions of the KOI data set inferred via machine learning. The values obtained from degraded solar data predicted on these quantities are shown for reference. \label{tab:results-kages-curr}}
\tablewidth{0pt}
\tablehead{\colhead{KIC} & \colhead{$\tau/$Gyr} & \colhead{X$_{\mathrm{c}}$} & \colhead{log g} & \colhead{L$/$L$_\odot$} & \colhead{R$/$R$_\odot$} & \colhead{Y$_{\mathrm{surf}}$}}
\startdata
 3425851 &  3.7 $\pm$ 0.76  & 0.14  $\pm$ 0.081  & 4.234 $\pm$ 0.0098 & 2.7  $\pm$ 0.16  & 1.36  $\pm$ 0.022  & 0.27  $\pm$ 0.026 \\
 3544595 &  6.7 $\pm$ 1.47  & 0.31  $\pm$ 0.078  & 4.46  $\pm$ 0.016  & 0.84 $\pm$ 0.068 & 0.94  $\pm$ 0.020  & 0.23  $\pm$ 0.023 \\
 3632418 &  3.0 $\pm$ 0.36  & 0.10  $\pm$ 0.039  & 4.020 $\pm$ 0.0076 & 5.2  $\pm$ 0.25  & 1.91  $\pm$ 0.031  & 0.24  $\pm$ 0.021 \\
 4141376 &  3.4 $\pm$ 0.67  & 0.38  $\pm$ 0.070  & 4.41  $\pm$ 0.011  & 1.42 $\pm$ 0.097 & 1.05  $\pm$ 0.019  & 0.24  $\pm$ 0.022 \\
 4143755 &  8.0 $\pm$ 0.80  & 0.07  $\pm$ 0.022  & 4.09  $\pm$ 0.013  & 2.3  $\pm$ 0.12  & 1.50  $\pm$ 0.029  & 0.17  $\pm$ 0.023 \\
 4349452 &  2.4 $\pm$ 0.78  & 0.4   $\pm$ 0.10   & 4.28  $\pm$ 0.012  & 2.5  $\pm$ 0.14  & 1.32  $\pm$ 0.022  & 0.22  $\pm$ 0.043 \\
 4914423 &  5.2 $\pm$ 0.58  & 0.06  $\pm$ 0.032  & 4.162 $\pm$ 0.0097 & 2.5  $\pm$ 0.16  & 1.50  $\pm$ 0.022  & 0.24  $\pm$ 0.023 \\
 5094751 &  5.3 $\pm$ 0.67  & 0.07  $\pm$ 0.039  & 4.209 $\pm$ 0.0082 & 2.2  $\pm$ 0.13  & 1.37  $\pm$ 0.017  & 0.23  $\pm$ 0.024 \\
 5866724 &  2.4 $\pm$ 0.96  & 0.4   $\pm$ 0.12   & 4.24  $\pm$ 0.017  & 2.7  $\pm$ 0.13  & 1.42  $\pm$ 0.022  & 0.23  $\pm$ 0.038 \\
 6196457 &  4.0 $\pm$ 0.73  & 0.18  $\pm$ 0.061  & 4.11  $\pm$ 0.022  & 3.3  $\pm$ 0.21  & 1.68  $\pm$ 0.041  & 0.24  $\pm$ 0.016 \\
 6278762 & 10.3 $\pm$ 0.96  & 0.35  $\pm$ 0.026  & 4.557 $\pm$ 0.0084 & 0.34 $\pm$ 0.022 & 0.761 $\pm$ 0.0061 & 0.19  $\pm$ 0.023 \\
 6521045 &  5.6 $\pm$ 0.370 & 0.027 $\pm$ 0.0097 & 4.122 $\pm$ 0.0055 & 2.7  $\pm$ 0.15  & 1.57  $\pm$ 0.025  & 0.22  $\pm$ 0.019 \\
 7670943 &  2.3 $\pm$ 0.59  & 0.32  $\pm$ 0.088  & 4.234 $\pm$ 0.0099 & 3.3  $\pm$ 0.23  & 1.44  $\pm$ 0.025  & 0.26  $\pm$ 0.029 \\
 8077137 &  4.4 $\pm$ 0.96  & 0.08  $\pm$ 0.052  & 4.08  $\pm$ 0.016  & 3.7  $\pm$ 0.24  & 1.68  $\pm$ 0.044  & 0.22  $\pm$ 0.031 \\
 8292840 &  3.4 $\pm$ 1.48  & 0.3   $\pm$ 0.14   & 4.25  $\pm$ 0.023  & 2.6  $\pm$ 0.20  & 1.34  $\pm$ 0.026  & 0.19  $\pm$ 0.049 \\
 8349582 &  6.7 $\pm$ 0.53  & 0.02  $\pm$ 0.012  & 4.16  $\pm$ 0.012  & 2.2  $\pm$ 0.12  & 1.52  $\pm$ 0.016  & 0.23  $\pm$ 0.015 \\
 8478994 &  4.6 $\pm$ 1.75  & 0.50  $\pm$ 0.055  & 4.55  $\pm$ 0.012  & 0.51 $\pm$ 0.036 & 0.79  $\pm$ 0.014  & 0.21  $\pm$ 0.022 \\
 8494142 &  2.8 $\pm$ 0.52  & 0.18  $\pm$ 0.067  & 4.06  $\pm$ 0.018  & 4.5  $\pm$ 0.32  & 1.84  $\pm$ 0.043  & 0.24  $\pm$ 0.029 \\
 8554498 &  3.7 $\pm$ 0.79  & 0.09  $\pm$ 0.060  & 4.04  $\pm$ 0.015  & 4.1  $\pm$ 0.20  & 1.86  $\pm$ 0.043  & 0.25  $\pm$ 0.018 \\
 8684730 &  3.0 $\pm$ 0.38  & 0.24  $\pm$ 0.065  & 4.06  $\pm$ 0.046  & 4.1  $\pm$ 0.53  & 1.9   $\pm$ 0.11   & 0.17  $\pm$ 0.040 \\
 8866102 &  1.9 $\pm$ 0.71  & 0.4   $\pm$ 0.11   & 4.27  $\pm$ 0.014  & 2.8  $\pm$ 0.16  & 1.36  $\pm$ 0.024  & 0.24  $\pm$ 0.039 \\
 9414417 &  3.1 $\pm$ 0.31  & 0.09  $\pm$ 0.030  & 4.016 $\pm$ 0.0058 & 5.0  $\pm$ 0.32  & 1.90  $\pm$ 0.032  & 0.21  $\pm$ 0.026 \\
 9592705 &  3.0 $\pm$ 0.38  & 0.05  $\pm$ 0.026  & 3.973 $\pm$ 0.0087 & 5.7  $\pm$ 0.37  & 2.06  $\pm$ 0.035  & 0.26  $\pm$ 0.015 \\
 9955598 &  7.0 $\pm$ 0.98  & 0.37  $\pm$ 0.035  & 4.494 $\pm$ 0.0061 & 0.66 $\pm$ 0.041 & 0.90  $\pm$ 0.013  & 0.25  $\pm$ 0.020 \\
10514430 &  6.5 $\pm$ 0.89  & 0.06  $\pm$ 0.022  & 4.08  $\pm$ 0.014  & 2.9  $\pm$ 0.17  & 1.62  $\pm$ 0.026  & 0.22  $\pm$ 0.021 \\
10586004 &  4.9 $\pm$ 1.39  & 0.12  $\pm$ 0.090  & 4.09  $\pm$ 0.041  & 3.1  $\pm$ 0.27  & 1.71  $\pm$ 0.070  & 0.24  $\pm$ 0.021 \\
10666592 &  2.0 $\pm$ 0.24  & 0.15  $\pm$ 0.036  & 4.020 $\pm$ 0.0066 & 5.7  $\pm$ 0.33  & 1.98  $\pm$ 0.018  & 0.29  $\pm$ 0.014 \\
10963065 &  4.4 $\pm$ 0.58  & 0.16  $\pm$ 0.054  & 4.292 $\pm$ 0.0070 & 2.0  $\pm$ 0.1   & 1.24  $\pm$ 0.015  & 0.22  $\pm$ 0.029 \\
11133306 &  4.1 $\pm$ 0.84  & 0.22  $\pm$ 0.079  & 4.319 $\pm$ 0.0096 & 1.7  $\pm$ 0.11  & 1.21  $\pm$ 0.019  & 0.22  $\pm$ 0.036 \\
11295426 &  6.2 $\pm$ 0.78  & 0.09  $\pm$ 0.036  & 4.283 $\pm$ 0.0059 & 1.65 $\pm$ 0.095 & 1.26  $\pm$ 0.016  & 0.24  $\pm$ 0.012 \\
11401755 &  5.6 $\pm$ 0.630 & 0.037 $\pm$ 0.0053 & 4.043 $\pm$ 0.0071 & 3.4  $\pm$ 0.19  & 1.69  $\pm$ 0.026  & 0.21  $\pm$ 0.026 \\
11807274 &  2.8 $\pm$ 1.05  & 0.3   $\pm$ 0.11   & 4.17  $\pm$ 0.024  & 3.5  $\pm$ 0.22  & 1.57  $\pm$ 0.038  & 0.22  $\pm$ 0.035 \\
11853905 &  5.7 $\pm$ 0.78  & 0.04  $\pm$ 0.020  & 4.11  $\pm$ 0.011  & 2.7  $\pm$ 0.16  & 1.62  $\pm$ 0.030  & 0.23  $\pm$ 0.022 \\
11904151 &  9.6 $\pm$ 1.43  & 0.08  $\pm$ 0.037  & 4.348 $\pm$ 0.0097 & 1.09 $\pm$ 0.06  & 1.07  $\pm$ 0.019  & 0.21  $\pm$ 0.026 \\
     Sun &  4.6 $\pm$ 0.16  & 0.36  $\pm$ 0.012  & 4.439 $\pm$ 0.0038 & 1.01 $\pm$ 0.041 & 1.000 $\pm$ 0.0066 & 0.245 $\pm$ 0.0076 
\enddata
\end{deluxetable*}

\section{Discussion}
The amount of time it takes to make predictions for a star using a trained random forest can be decomposed into two parts: the amount of time it takes to calculate perturbations to the observations of the star \mb{(see Section \ref{sec:uncertainties})}, and the amount of time it takes to make a prediction on each perturbed set of observations. Hence we have
\begin{equation}
    t = n(t_p + t_r)
\end{equation}
where $t$ is the total time, $n$ is the number of perturbations, $t_p$ is the time it takes to perform a single perturbation, and $t_r$ is the random forest regression time. We typically see times of $t_p = (7.9 \pm 0.7) \cdot 10^{-3}$ $(\si{\s})$ and $t_r = (1.8 \pm 0.4) \cdot 10^{-5}$ $(\si{\s})$. We chose a conservative $n=10\,000$ for the results presented here, which results in a time of around a minute per star. Since each star can be processed independently and in parallel, a computing cluster could feasibly process a catalog containing millions of objects in less than a day. Since \mb{$t_r \ll t_p$}, the calculation depends almost entirely on the time it takes to perturb the observations.\footnote{Our perturbation code uses an interpreted language (R), so if needed, there is still room for speed-up.} There is also the one-time cost of training the random forest, which takes less than a minute and can be reused without retraining on every star with the same information. It does need to be retrained if one wants to consider a different combination of input or output parameters. 

There is a one-time cost of generating the matrix of training data. We ran our simulation generation scheme for a week on our computing cluster and obtained 5\,325 evolutionary tracks with 64 models per track, \mb{which resulted in a} 123 MB \mb{matrix of stellar models}. This is at least an order of magnitude fewer models than the amount that other methods use. Furthermore, this is in general more tracks than is needed by our method: we showed in Figure \ref{fig:evaluation-tracks} that for most parameters---most notably age, mass, luminosity, radius, initial metallicity, and core hydrogen abundance---one needs only a fraction of the models that we generated in order to obtain good predictive accuracies. Finally, unless one wants to consider a different range of parameters or different input physics, this matrix would not need to be calculated again; a random forest trained on this matrix can be re-used for all future stars that are observed. Of course, our method would still work if trained using a different matrix of models, and our grid should work with other grid-based modelling methods. 

Previously, \citet{pulone1997age} developed a neural network for predicting stellar age based on the star's position in the Hertzsprung-Russell diagram. More recently, \citet{2016arXiv160200902V} have worked on incorporating seismic information into that analysis as we have done here. Our method provides several advantages over these approaches. Firstly, the random forests that we use perform constrained regression, meaning that the values we predict for quantities like age and mass will always be non-negative and within the bounds of the training data, which is not true of the neural networks-based approach that they take. Secondly, using \emph{averaged} frequency separations allows us to make predictions without need for concern over which radial orders were observed. Thirdly, we have shown that our random forests are very fast to train, and can be retrained in only seconds for stars that are missing observational constraints such as luminosities. In contrast, deep neural networks are computationally intensive to train, taking days or even weeks to converge depending on the breadth of network topologies considered in the cross-validation. Finally, our grid is varied in six initial parameters---M, Y$_0$, Z$_0$, $\alpha_{\text{MLT}}$, $\alpha_{\text{ov}}$, and D, which allows our method to explore a wide range of stellar model parameters.

\section{Conclusions}
Here we have considered the constrained multiple-regression problem of inferring fundamental stellar parameters from observations. We created a grid of evolutionary tracks varied in mass, chemical composition, mixing length parameter, overshooting coefficient, and diffusion \mb{multiplication} factor. We evolved each track in time along the main sequence and collected \mb{observable quantities} such as effective temperatures and metallicities as well as global statistics on the modes of oscillations from models along each evolutionary path. We used this matrix of \mb{stellar models} to train a machine learning algorithm to be able to discern the patterns that relate observations to \mb{fundamental stellar parameters}. We then applied this method to hare-and-hound exercise data, the Sun, 16 Cyg A and B, and 34 planet-hosting candidates that have been observed by \emph{Kepler} and rapidly obtained precise initial conditions and current-age values of these stars. 
Remarkably, we were able to empirically determine the value of the diffusion \mb{multiplication} factor and hence the efficiency of diffusion required to reproduce the observations instead of inhibiting it \emph{ad hoc}. A larger sample size \mb{will} better constrain the diffusion \mb{multiplication} factor and determine what other variables are relevant in its parametrization. \mb{This is work in progress.}

The method presented here has many advantages over existing approaches. First, random forests can be trained and used in only seconds and hence provide substantial speed-ups over other methods. Observations of a star simply need to be fed through the forest---akin to plugging numbers into an equation---and do not need to be subjected to expensive iterative optimization procedures. 
Secondly, random forests perform non-linear and non-parametric regression, which means that the method can use orders-of-magnitude fewer models for the same level of precision, while additionally attaining a more rigorous appraisal of uncertainties for the predicted quantities. 
Thirdly, our method allows us to investigate wide ranges and combinations of stellar parameters. 
And finally, the method presented here provides the opportunity to extract insights from the statistical regression that is being performed, which is achieved by examining the relationships in stellar physics that the machine learns by analyzing simulation data. This contrasts the blind optimization processes of other methods that provide an answer but do not indicate the elements that were important in doing so. 

We note that the predicted quantities reflect a set of choices in stellar physics. Although such biases are impossible to propagate, varying model parameters that are usually kept fixed---such as the mixing length parameter, diffusion \mb{multiplication} factor, and overshooting coefficient---takes us a step in the right direction. Furthermore, the fact that quantities such as stellar radii and luminosities---quantities that have been measured accurately, not just precisely---can be reproduced both precisely and accuractely by this method, gives a degree of confidence in its efficacy. 

The method we have presented here is currently only applicable to main-sequence stars. We intend to extend this study to later stages of evolution.

\acknowledgments The research leading to the presented results has received funding from the European Research Council under the European Community's Seventh Framework Programme (FP7/2007-2013) / ERC grant agreement no 338251 (StellarAges). This research was undertaken in the context of the International Max Planck Research School \mb{for Solar System Research}. S.B.\ acknowledges partial support from NSF grant AST-1514676 and NASA grant NNX13AE70G. W.H.B.\ acknowledges research funding by Deutsche Forschungsgemeinschaft (DFG) under grant SFB 963/1 ``Astrophysical flow instabilities and turbulence'' (Project A18).

\software Analysis in this manuscript was performed with \mb{python 3.5.1} libraries scikit-learn \mb{0.17.1} \citep{scikit-learn}, NumPy \mb{1.10.4} \citep{van2011numpy}, and pandas \mb{0.17.1} \citep{mckinney2010data} as well as \mb{R 3.2.3} \citep{R} and the R libraries magicaxis \mb{1.9.4} \citep{magicaxis}, RColorBrewer \mb{1.1-2} \citep{RColorBrewer}, parallelMap \mb{1.3} \citep{parallelMap}, data.table \mb{1.9.6} \citep{data.table}, lpSolve \mb{5.6.13} \citep{lpSolve}, ggplot2 \mb{2.1.0} \citep{ggplot2}, GGally \mb{1.0.1} \citep{GGally}, scales \mb{0.3.0} \citep{scales}, deming \mb{1.0-1} \citep{deming}, and matrixStats \mb{0.50.1} \citep{matrixStats}.

\appendix

\section{Model selection}
\label{sec:selection}
To prevent statistical bias towards the evolutionary tracks that generate the most models, i.e.\ the ones that require the most careful calculations and therefore use smaller time-steps, or those that live on the main sequence for a longer amount of time; we select $n=64$ models from each evolutionary track such that the models are as evenly-spaced in core hydrogen abundance as possible. \mb{We chose 64 because it is a power of two, which thus allows us to successively omit every other model when testing our regression routine and still maintain regular spacings.}

Starting from the original vector of length $m$ of core hydrogen abundances $\vec X$, we find the subset of length $m$ that is closest to the optimal spacing $\vec B$, where
\begin{equation}
  \vec B \equiv \left[
    X_T, 
    \ldots, 
    \frac{(m-i)\cdot X_T + X_Z}{m-1}, 
    \ldots, 
    X_Z
  \right]
\end{equation}
with $X_Z$ being the core hydrogen abundance at ZAMS and $X_T$ being that at TAMS. To obtain the closest possible vector to $\vec B$ from our data $\vec X$, we solve a transportation problem using integer optimization \citep{23145595}. First we set up a cost matrix $\boldsymbol{C}$ consisting of absolute differences between the original abundances $\vec X$ and the ideal abundances $\vec B$:
\begin{equation} 
  \boldsymbol{C} \equiv \left[
  \begin{array}{cccc}
    \abs{B_1-X_1} & \abs{B_1-X_2} & \dots & \abs{B_1-X_n} \\
    \abs{B_2-X_1} & \abs{B_2-X_2} & \dots & \abs{B_2-X_n} \\
    \vdots & \vdots & \ddots & \vdots \\
    \abs{B_m-X_1} & \abs{B_m-X_2} & \dots & \abs{B_m-X_n}
  \end{array} \right].
\end{equation}
We then require that exactly $m$ values are selected from $\vec X$, and that each value is selected no more than one time. Simply selecting the closest data point to each ideally-separated point will not work because this could result in the same point being selected twice; and selecting the second closest point in that situation does not remedy it because a different result could be obtained if the points were processed in a different order. 

\begin{figure*}
    \centering
    \includegraphics[width=\linewidth, keepaspectratio]{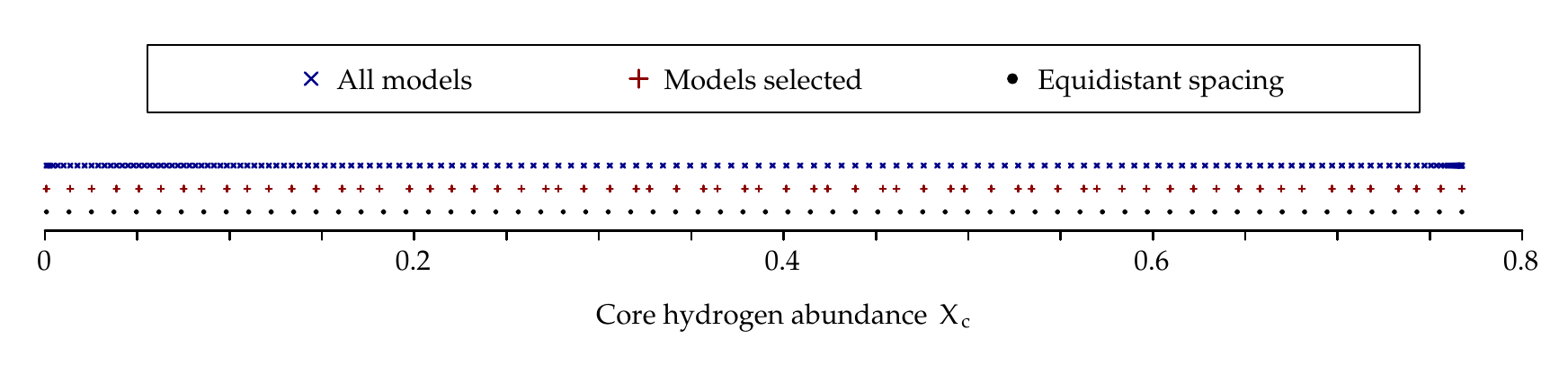}
    \caption{ A visaulization of the model selection process performed on each evolutionary track in order to obtain the same number of models from each track. The blue crosses show all of the models along the evolutionary track as they vary from ZAMS to TAMS in core hydrogen abundance and the red crosses show the models selected from this track. The models were chosen via linear transport such that they satisfy Equation \ref{eq:optimal-spacing}. For reference, an equidistant spacing is shown with black points. \vspace*{5mm}
    \label{fig:nearly-even} }%
\end{figure*}

We denote the optimal solution matrix by $\hat{\boldsymbol{S}}$, and find it by minimising the cost matrix subject to the following constraints:
\begin{align}
  \hat{\boldsymbol{S}} = \underset{\boldsymbol S}{\arg\min} \; & \sum_{ij} S_{ij} C_{ij} \notag\\
  \text{subject to } & \sum_j S_{ij} \leq 1 \; \text{ for all } i=1\ldots N \notag\\
  \text{and } & \sum_i S_{ij} = 1 \; \text{ for all } j=1\ldots M.
  \label{eq:optimal-spacing}
\end{align}
The indices of $\vec X$ that are most near to being equidistantly-spaced are then found by looking at which columns of $\hat{\boldsymbol S}$ contain ones. The solution is visualized in Figure \ref{fig:nearly-even}.

\section{Initial grid strategy}
\label{sec:grid}
The initial conditions of a stellar model can be viewed as a \mb{six-dimensional hyperrectangle} with dimensions M, Y$_0$, Z$_0$, $\alpha_{\text{MLT}}$, $\alpha_{\text{ov}}$, and D. In order to vary all of these parameters simultaneously and fill the \mb{hyperrectangle} as quickly as possible, we construct a grid of initial conditions following a quasi-random point generation scheme. This is in contrast to linear or random point generation schemes, over which it has several advantages. 

A linear grid subdivides all dimensions in which initial quantities can vary into equal parts and creates a track of models for every combination of these subdivisions. Although in the limit such a strategy will fill the \mb{hyperrectangle} of initial conditions, it does so very slowly. It is furthermore suboptimal in the sense that linear grids maximize redundant information, as each varied quantity is tried with the exact same values of all other parameters that have been considered already. In a high-dimensional setting, if any of the parameters are irrelevant to the task of the computation, then the majority of the tracks in a linear grid will not contribute any new information.

A refinement on this approach is to create a grid of models with randomly varied initial conditions. Such a strategy fills the space more rapidly, and furthermore solves the problem of redundant information. However, this approach suffers from a different problem: since the points are generated at random, they tend to ``clump up'' at random as well. This results in random gaps in the parameter space, which are obviously undesirable. 

\begin{figure*}
    \centering
    \includegraphics[width=0.32\linewidth,keepaspectratio]{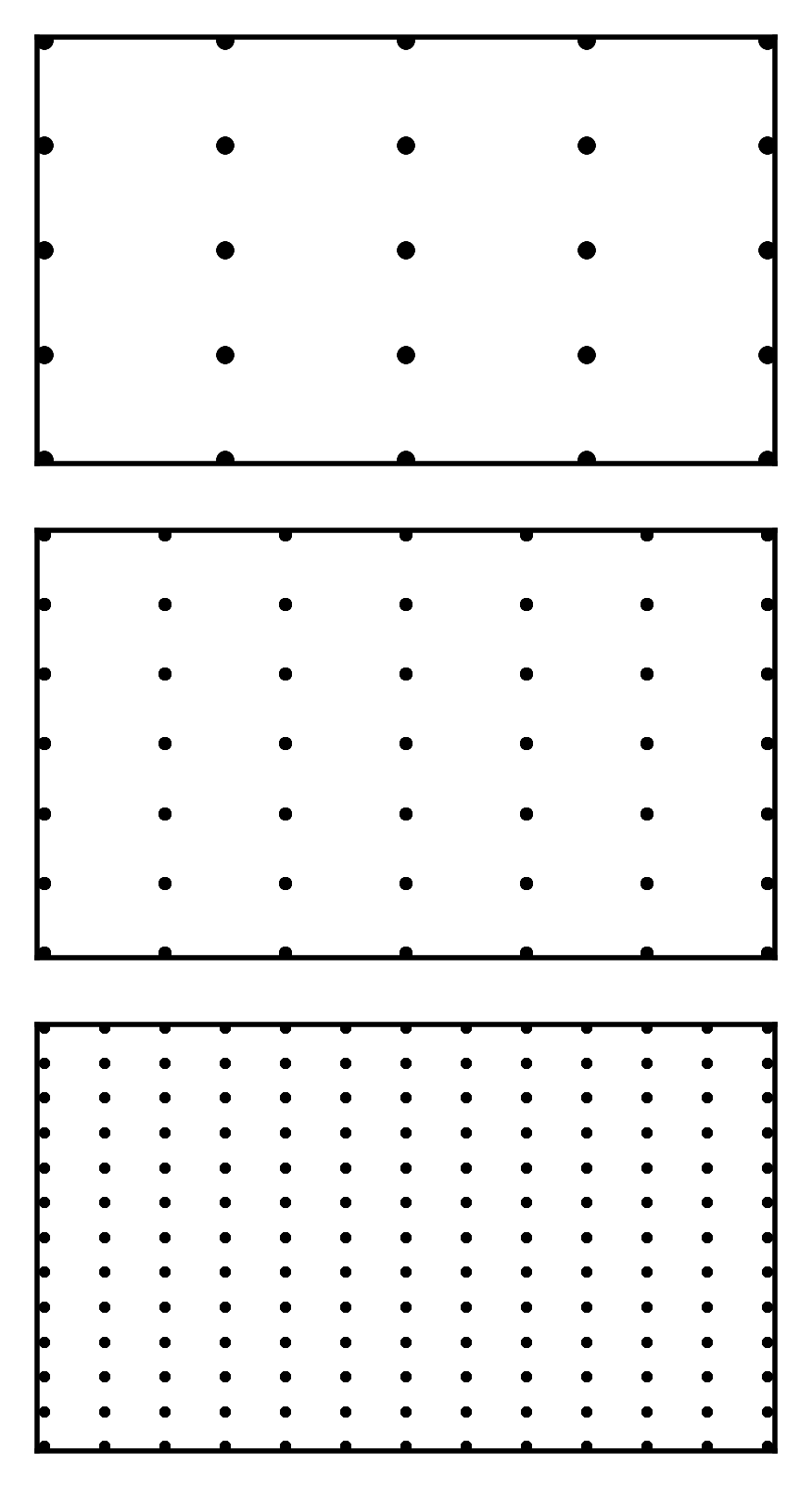}\hfill
    \includegraphics[width=0.32\linewidth,keepaspectratio]{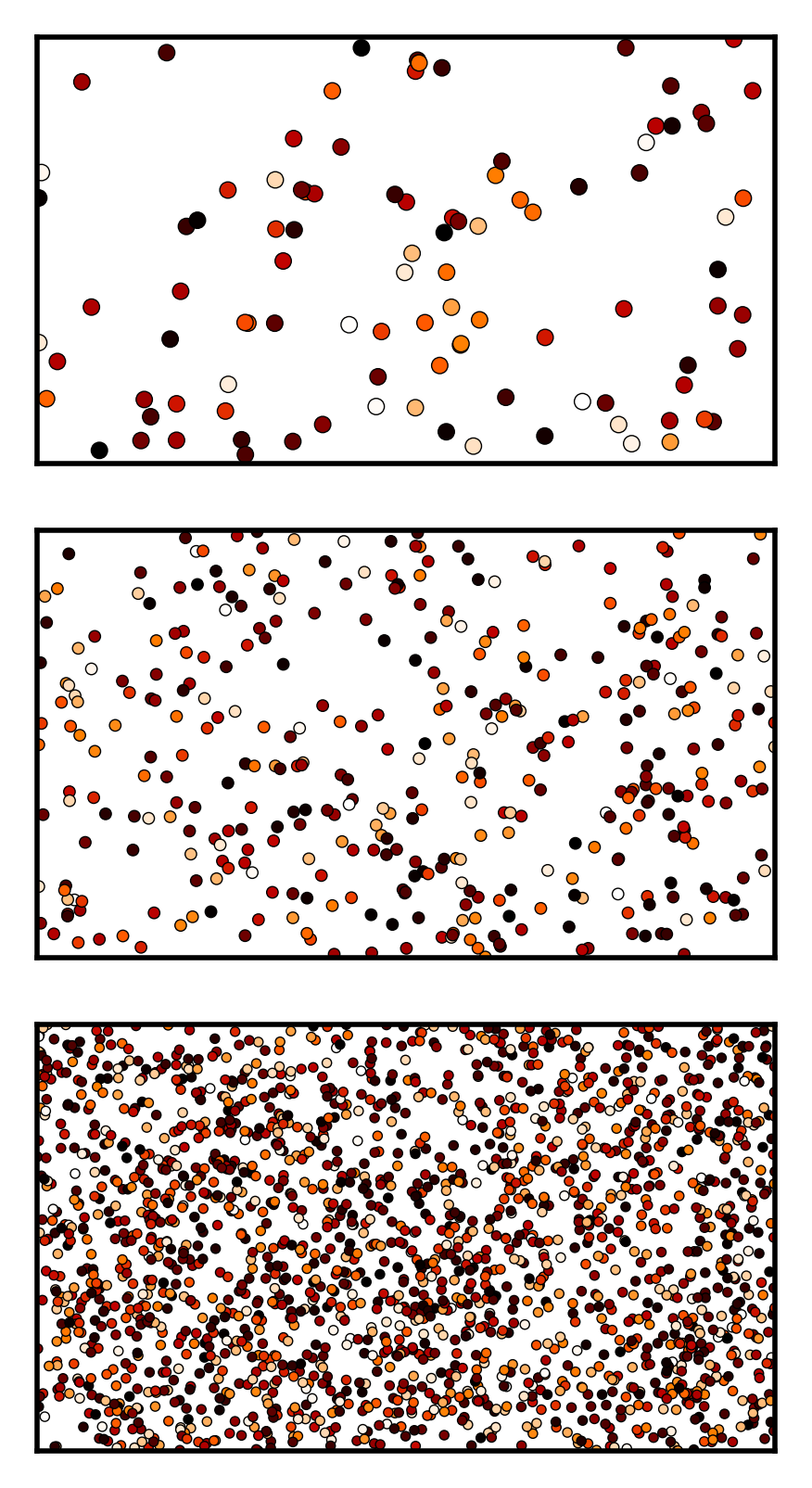}\hfill
    \includegraphics[width=0.32\linewidth,keepaspectratio]{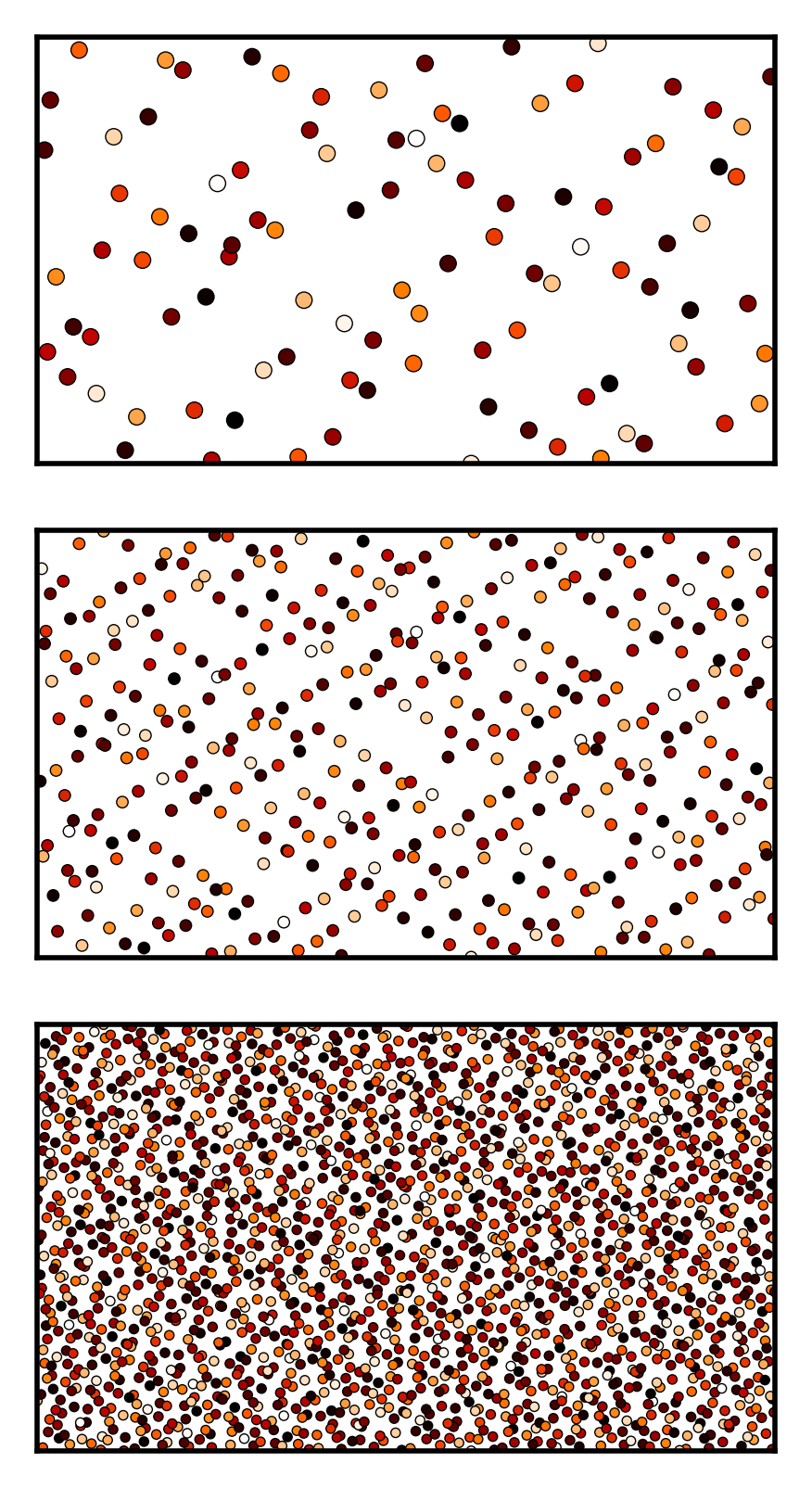}\\
    \parbox{0.32\linewidth}{\centering Linear}\hfill
    \parbox{0.32\linewidth}{\centering Random}\hfill
    \parbox{0.32\linewidth}{\centering Quasi-random}
    \caption{Results of different methods for generating multidimensional grids portrayed via a unit cube projected onto a unit square. Linear (left), random (middle), and quasi-random (right) grids are generated in three dimensions, with color depicting the third dimension, i.e., the distance between the reader and the screen. From top to bottom, all three methods are shown with 100, 400, and 2000 points generated, respectively. }%
    \label{fig:grids}
\end{figure*}

Therefore, in order to select points that do not stack, do not clump, and also fill the space as rapidly as possible, we generate Sobol numbers \citep{sobol1967distribution} in the unit 6-cube and map them to the parameter ranges of each quantity that we want to vary. Sobol numbers are a sequence of $m$-dimensional vectors $x_1 \ldots x_n$ in the unit hypercube $I^m$ constructed such that the integral of a real function $f$ in that space is equivalent in the limit to that function evaluated on those numbers, that is,
\begin{equation}
    \int_{I^m} f = \lim_{n \to \infty} \frac{1}{n}\sum_{i=1}^n f(x_i)
\end{equation}
with the sequence being chosen such that the convergence is achieved as quickly as possible. By doing this, we both minimize redundant information and furthermore sample the hyperspace of possible stars as uniformly as possible. Figure \ref{fig:grids} visualizes the different methods of generating multidimensional grids: linear, random, and the quasi-random strategy that we took. This method applied to initial model conditions was shown in Figure \ref{fig:inputs} with 1- and 2D projection plots of the evolutionary tracks generated for our grid. 

\section{Adaptive Remeshing}
\label{sec:remeshing}

\begin{figure}
    \centering
    \includegraphics[width=\colwidth,keepaspectratio]{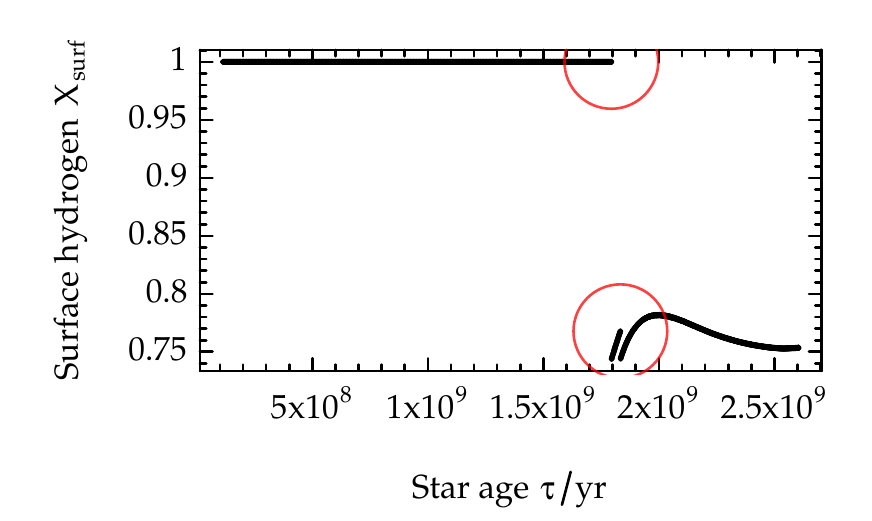}\\
    \includegraphics[width=\colwidth,keepaspectratio]{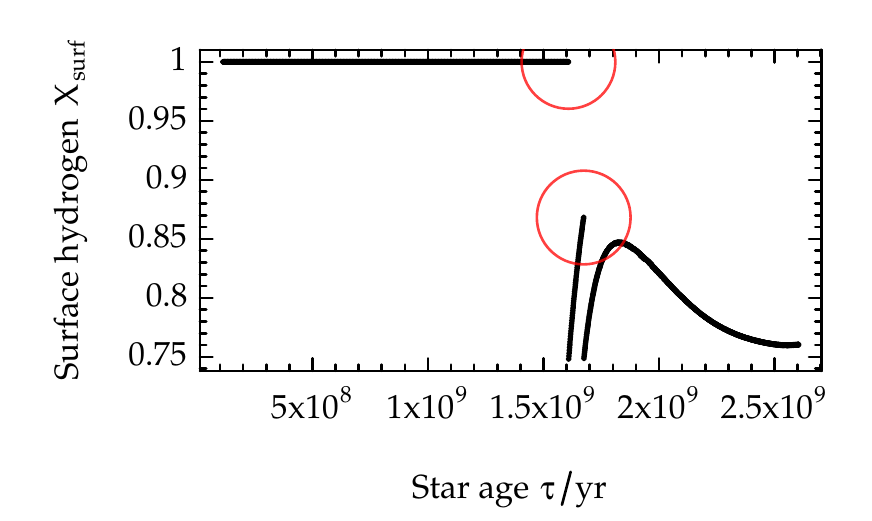}\\
    \includegraphics[width=\colwidth,keepaspectratio]{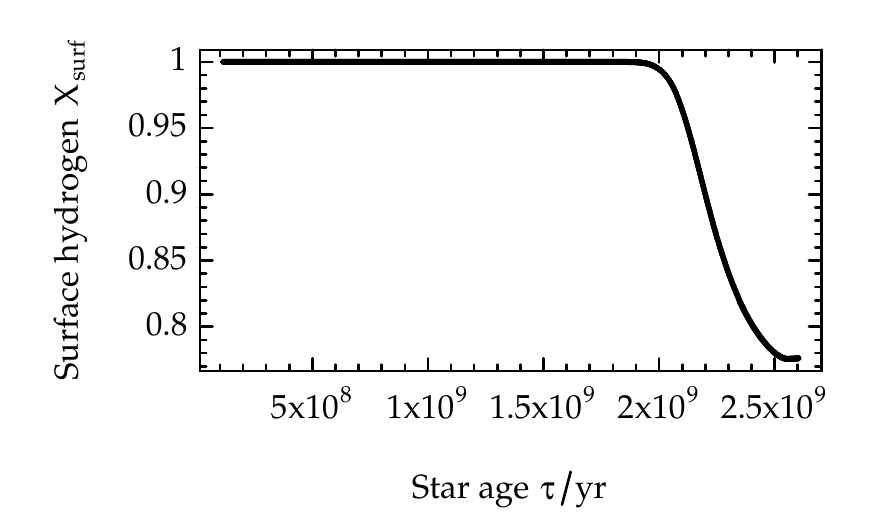}\\
    \caption{Three iterations of surface abundance discontinuity detection and iterative remeshing for an evolutionary track. The detected discontinuities are encircled in red. The third iteration has no discontinuities and so this track is considered to have converged. \vspace*{5mm} \label{fig:discontinuity} }
\end{figure}

When performing element diffusion calculations in MESA, the surface abundance of each isotope is considered as an average over the outermost cells of the model. The number of outer cells $N$ is chosen such that the mass of the surface is more than ten times the mass of the $(N+1)^{\text{th}}$ cell. Occasionally, this approach can lead to a situation where surface abundances change dramatically and discontinuously in a single time-step. These abundance discontinuities then propagate as discontinuities in effective temperatures, surface gravities, and radii. An example of such a difficulty \mb{is shown} in Figure \ref{fig:discontinuity}. 

Instead of being a physical reality, these effects arise only when there is insufficient mesh resolution in the outermost layers of the model. We therefore seek to detect these cases and re-run any such evolutionary track using a finer mesh resolution. We consider a track an outlier if its surface hydrogen abundance changes by more than 1\% in a single time-step. We iteratively re-run any track with outliers detected using a finer mesh resolution, and, if necessary, smaller time-steps, until convergence is reached. The process and a resolved track can also be seen in Figure \ref{fig:discontinuity}. 

\begin{figure*}
    \centering
    \includegraphics[width=0.5\linewidth,keepaspectratio]{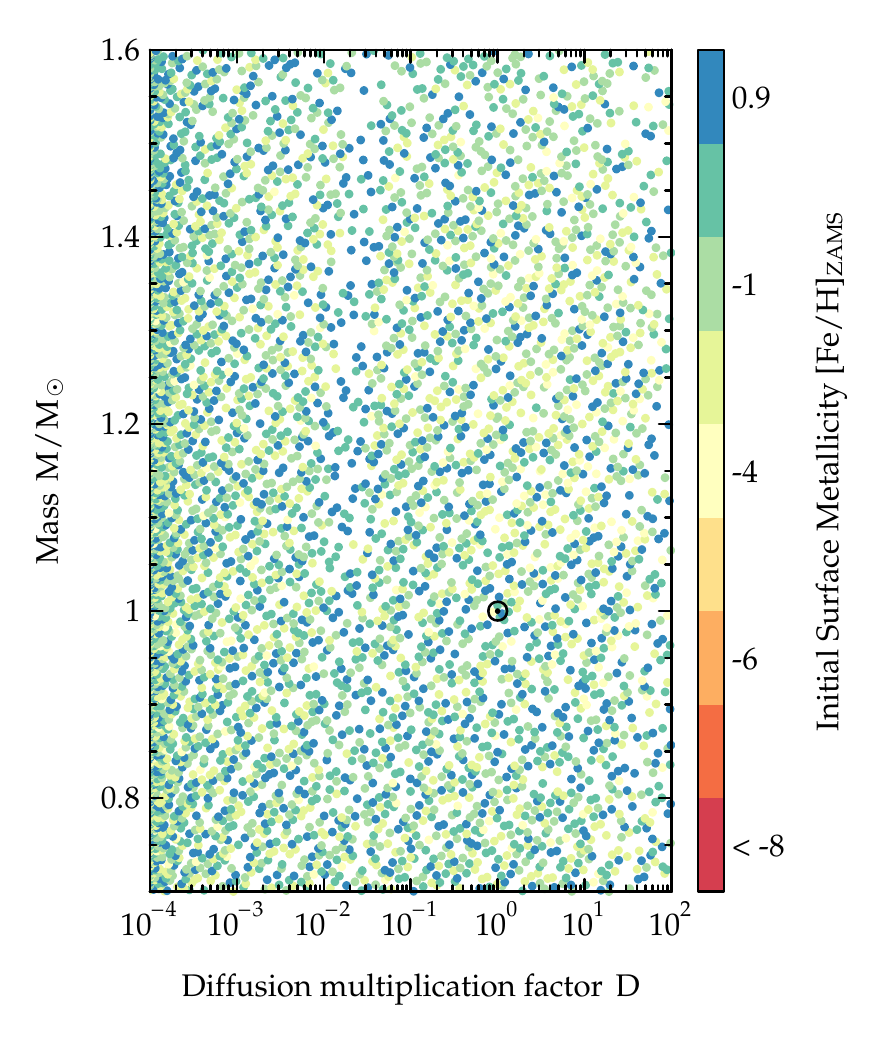}\hfill
    \includegraphics[width=0.5\linewidth,keepaspectratio]{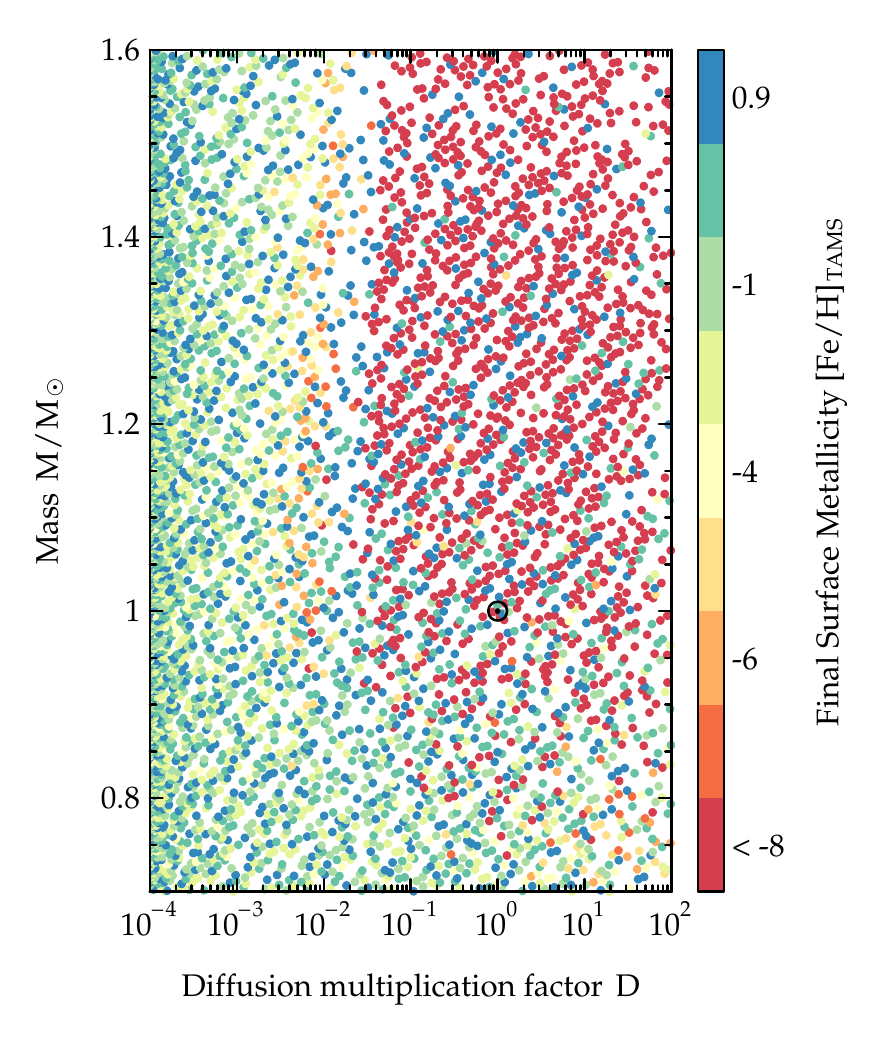}
    \caption{Stellar mass as a function of diffusion \mb{multiplication} factor colored by initial surface metallicity (left) and final surface metallicity (right). A ridge of \mb{missing points indicating} unconverged evolutionary tracks can be seen around a diffusion \mb{multiplication} factor of 0.01. Beyond this ridge, tracks that were initially metal-poor end their main-sequence lives with all of their metals drained from their surfaces. \label{fig:diffusion-gap} }
\end{figure*}

Some tracks still do not converge without surface abundance discontinuities despite the fineness of the mesh or the brevity of the time-steps, and are therefore not included in our study. These troublesome evolutionary tracks seem to be located only in a thin ridge of models having sufficiently high stellar mass (M $>$ 1), a deficit of initial metals (Z $<$ 0.001) and a specific inefficiency of diffusion (D $\simeq$ 0.01). A visualization of this \mb{is shown} in Figure \ref{fig:diffusion-gap}.

\section{Evaluating the regressor}
\label{sec:evaluation}
In training the random forest regressor, we must determine how many evolutionary tracks $N$ to include, how many models $M$ to extract from each evolutionary track, and how many trees $T$ to use when growing the forest. As such it is useful to define measures of gauging the accuracy of the random forest so that we may evaluate it with different combinations of these parameters. 

By far the most common way of measuring the quality of a random forest regressor is its so-called ``out-of-bag'' (OOB) score \citep[see e.g.\ section 3.1 of][]{breiman2001random}. While each tree is trained on only a subset (or ``bag'') of the stellar models, all trees are tested on all of the models that they did not see. This provides an accuracy score representing how well the forest will perform when predicting on observations that it has not seen yet. We can then use the scores defined in Section \ref{sec:uncertainties} to calculate OOB scores. 

However, such an approach to scoring is too optimistic in this scenario. Since a tree can get models from every simulation, predicting the \mb{parameters} of a model when the tree has been trained on one of that model's neighbors leads to an artificially inflated OOB score. This is especially the case for quantities like stellar mass, which do not change along the main sequence. A tree that has witnessed neighbors on either side of the model being predicted will have no error when predicting that model's mass, and hence the score will seem artificially better than it should be. 

Therefore, we opt instead to build validation sets containing entire tracks that are left out from the training of the random forest. We omit models and tracks in powers of two so that we may roughly maintain the regular spacing that we have established in our grid of models (refer back to Appendices \ref{sec:grid} and \ref{sec:selection} for details). 

\begin{figure*}
    \centering 
    \includegraphics[width=0.66\linewidth,keepaspectratio]{figs/evaluation/legend.png}\\
    \includegraphics[width=0.5\linewidth,keepaspectratio]{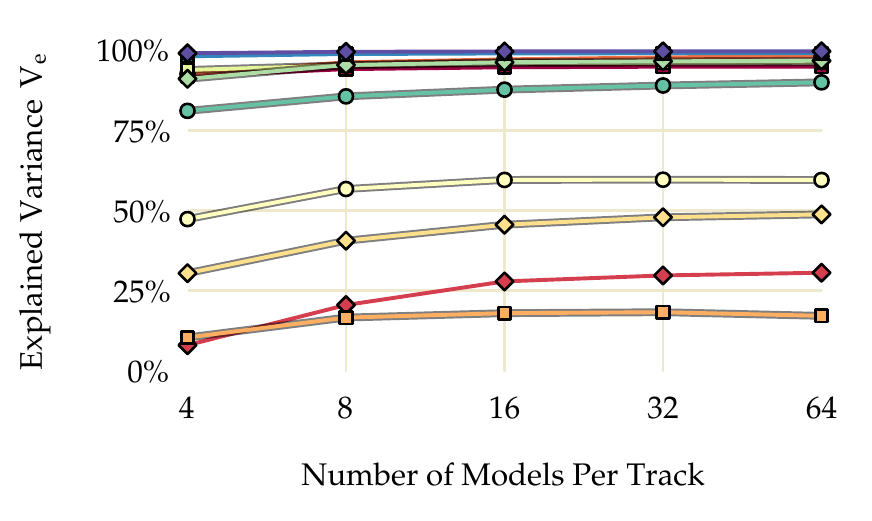}%
    \includegraphics[width=0.5\linewidth,keepaspectratio]{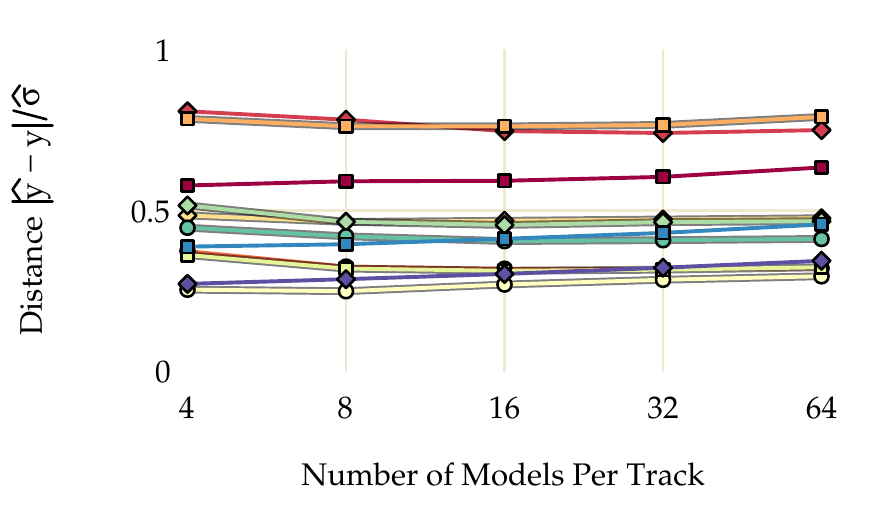}\\
    \includegraphics[width=0.5\linewidth,keepaspectratio]{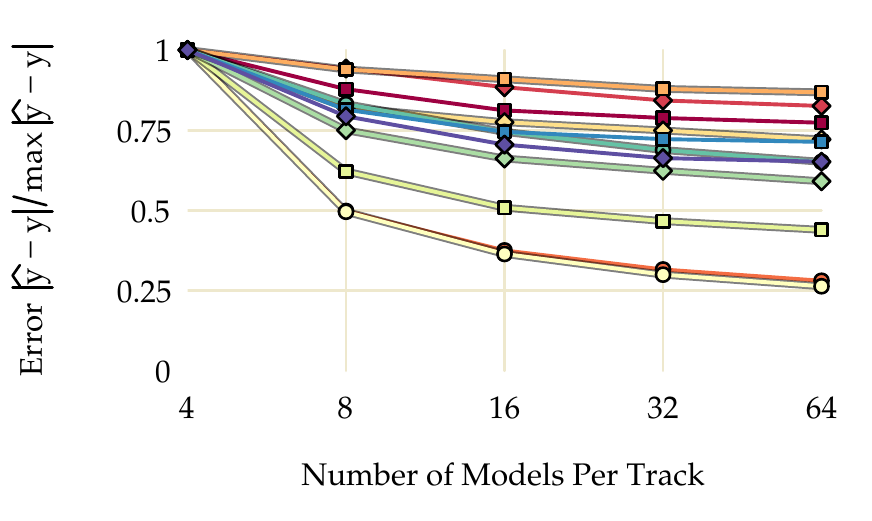}%
    \includegraphics[width=0.5\linewidth,keepaspectratio]{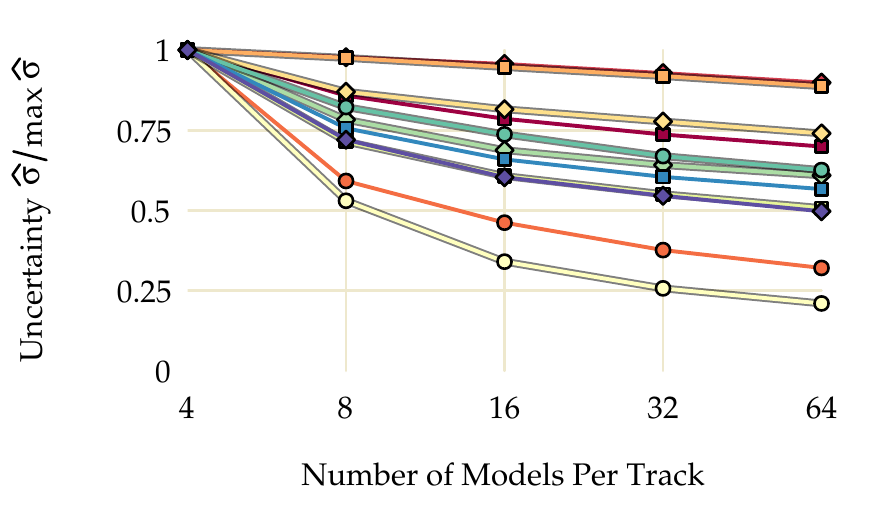}\\
    \caption{
    Explained variance (top left), accuracy per precision distance (top right), normalized absolute error (bottom left), and normalized standard deviation of predictions (bottom right) for each stellar parameter as a function of the number of models per evolutionary track.} 
    \label{fig:app-evaluation-models}
\end{figure*}

\begin{figure*}
    \centering
    \includegraphics[width=0.66\linewidth,keepaspectratio]{figs/evaluation/legend.png}\\
    \includegraphics[width=0.5\linewidth,keepaspectratio]{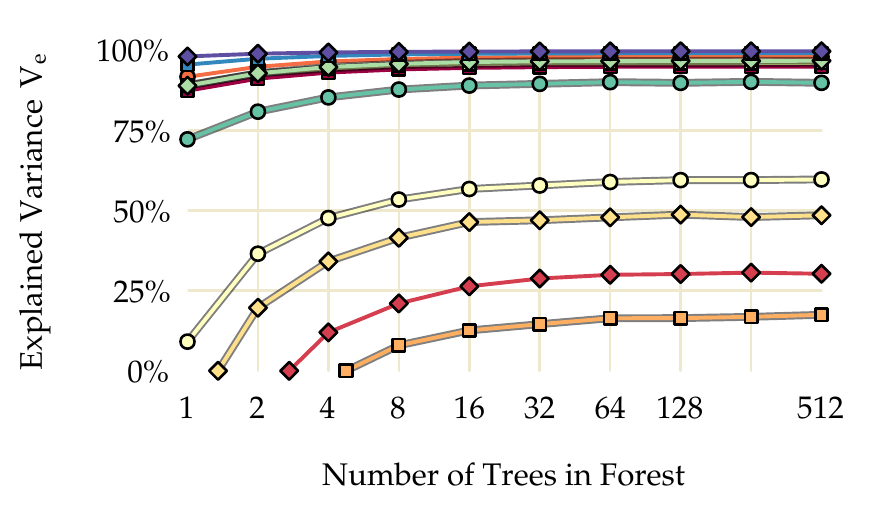}%
    \includegraphics[width=0.5\linewidth,keepaspectratio]{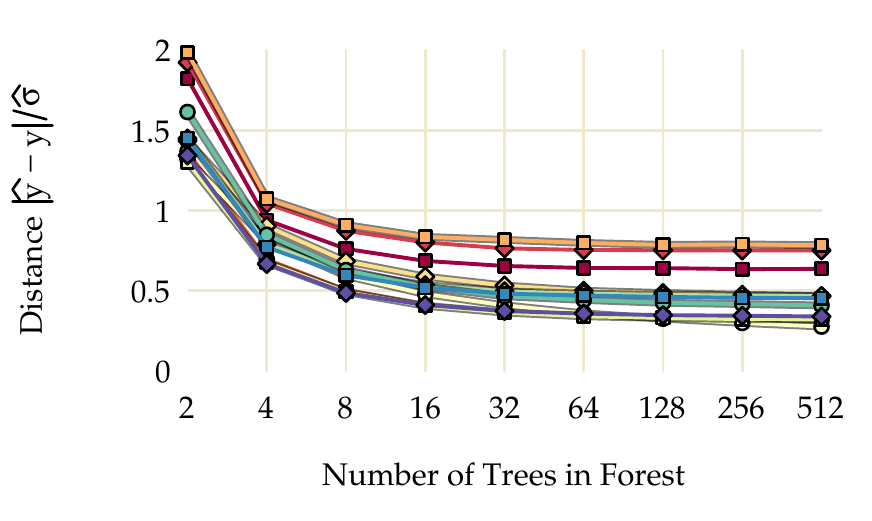}\\
    \includegraphics[width=0.5\linewidth,keepaspectratio]{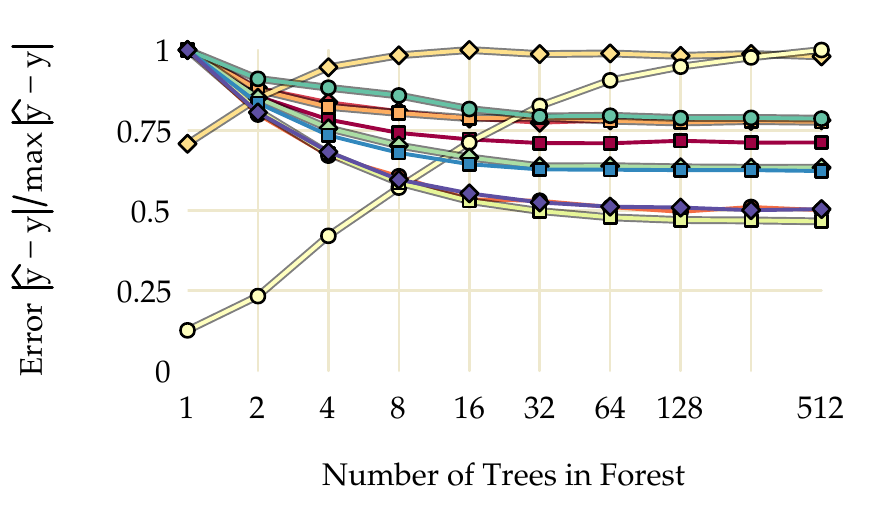}%
    \includegraphics[width=0.5\linewidth,keepaspectratio]{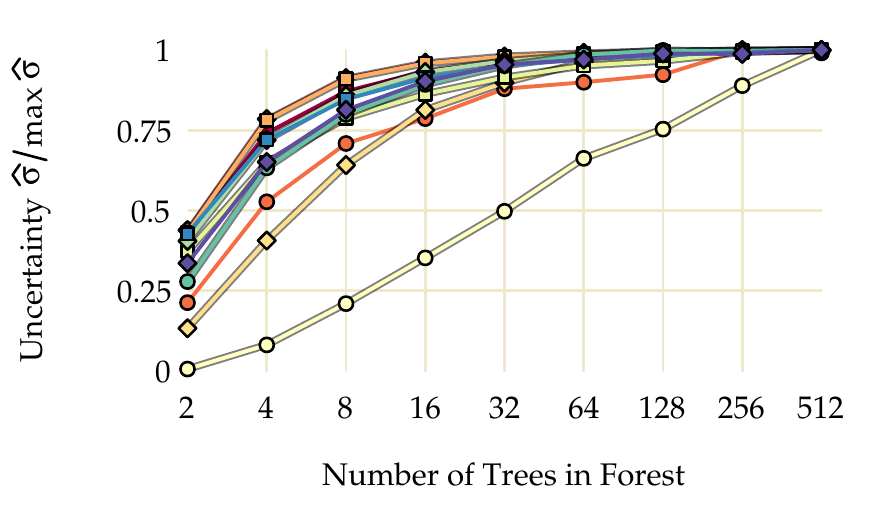}\\
    \caption{Explained variance (top left), accuracy per precision distance (top right), normalized absolute error (bottom left), and normalized model uncertainty (bottom right) for each stellar parameter as a function of the number of trees used in training the random forest. 
    \label{fig:app-evaluation-trees}} 
\end{figure*}

We have already shown in Figure \ref{fig:evaluation-tracks} these cross-validated scores as a function of the number of evolutionary tracks. Figure \ref{fig:app-evaluation-models} now shows these scores as a function of the number of models obtained from each evolutionary track, and Figure \ref{fig:app-evaluation-trees} shows them as a function of the number of trees in the forest. Naturally, $\hat\sigma$ increases with the number of trees, but this is not a mark against having more trees: this score is trivially minimal when there is only one tree, as that tree must agree with itself! We find that although more is better for all quantities,  there is not much improvement after about $T=32$ and $M=16$. It is also interesting to note that the predictions do not suffer very much from using only four models per track, which results in a random forest trained on only a few thousand models. 

\section{Hare-and-Hound}
\label{sec:hare-and-hound}
Table \ref{tab:hnh-true} lists the true values of the hare-and-hound exercise performed here, and Table \ref{tab:hnh-perturb} lists the perturbed inputs that were supplied to the machine learning algorithm. 

\begin{deluxetable*}{ccccccccccc}
\tabletypesize{\scriptsize}
\tablecaption{True values for the hare-and-hound exercise. \label{tab:hnh-true}}
\tablewidth{0pt}
\tablehead{\colhead{Model} & \colhead{R/R$_\odot$} & \colhead{M/M$_\odot$} & \colhead{$\tau$} & \colhead{T$_{\text{eff}}$} & \colhead{L/L$_\odot$} & \colhead{[Fe/H]} & \colhead{Y$_0$} & \colhead{$\nu_{\max}$} & \colhead{$\alpha_{\text{ov}}$} & \colhead{D}}
\startdata
0 & 1.705 & 1.303 & 3.725 & 6297.96 & 4.11 & 0.03 & 0.2520 & 1313.67 & No & No \\
1 & 1.388 & 1.279 & 2.608 & 5861.38 & 2.04 & 0.26 & 0.2577 & 2020.34 & No & No \\
2 & 1.068 & 0.951 & 6.587 & 5876.25 & 1.22 & 0.04 & 0.3057 & 2534.29 & No & No \\
3 & 1.126 & 1.066 & 2.242 & 6453.57 & 1.98 & -0.36 & 0.2678 & 2429.83 & No & No \\
4 & 1.497 & 1.406 & 1.202 & 6506.26 & 3.61 & 0.14 & 0.2629 & 1808.52 & No & No \\
5 & 1.331 & 1.163 & 4.979 & 6081.35 & 2.18 & 0.03 & 0.2499 & 1955.72 & No & No \\
6 & 0.953 & 0.983 & 2.757 & 5721.37 & 0.87 & -0.06 & 0.2683 & 3345.56 & No & No \\
7 & 1.137 & 1.101 & 2.205 & 6378.23 & 1.92 & -0.31 & 0.2504 & 2483.83 & No & No \\
8 & 1.696 & 1.333 & 2.792 & 6382.22 & 4.29 & -0.07 & 0.2555 & 1348.83 & No & No \\
9 & 0.810 & 0.769 & 9.705 & 5919.70 & 0.72 & -0.83 & 0.2493 & 3563.09 & No & No \\
10 & 1.399 & 1.164 & 6.263 & 5916.71 & 2.15 & 0.00 & 0.2480 & 1799.10 & Yes & Yes \\
11 & 1.233 & 1.158 & 2.176 & 6228.02 & 2.05 & 0.11 & 0.2796 & 2247.53 & Yes & Yes 
\enddata
\end{deluxetable*}

\begin{deluxetable*}{ccccc}
\tabletypesize{\scriptsize}
\tablecaption{Supplied (perturbed) inputs for the hare-and-hound exercise. \label{tab:hnh-perturb}}
\tablewidth{0pt}
\tablehead{\colhead{Model} & \colhead{T$_{\text{eff}}$} & \colhead{L/L$_\odot$} & \colhead{[Fe/H]} & \colhead{$\nu_{\max}$}}
\startdata
 0 & 6237 $\pm$ 85 & 4.2 $\pm$ 0.12 & -0.03 $\pm$ 0.09 & 1398 $\pm$  66 \\
 1 & 5806 $\pm$ 85 & 2.1 $\pm$ 0.06 &  0.16 $\pm$ 0.09 & 2030 $\pm$ 100 \\
 2 & 5885 $\pm$ 85 & 1.2 $\pm$ 0.04 & -0.05 $\pm$ 0.09 & 2630 $\pm$ 127 \\
 3 & 6422 $\pm$ 85 & 2.0 $\pm$ 0.06 & -0.36 $\pm$ 0.09 & 2480 $\pm$ 124 \\
 4 & 6526 $\pm$ 85 & 3.7 $\pm$ 0.11 &  0.14 $\pm$ 0.09 & 1752 $\pm$  89 \\
 5 & 6118 $\pm$ 85 & 2.2 $\pm$ 0.06 &  0.04 $\pm$ 0.09 & 1890 $\pm$ 101 \\
 6 & 5741 $\pm$ 85 & 0.8 $\pm$ 0.03 &  0.06 $\pm$ 0.09 & 3490 $\pm$ 165 \\
 7 & 6289 $\pm$ 85 & 2.0 $\pm$ 0.06 & -0.28 $\pm$ 0.09 & 2440 $\pm$ 124 \\
 8 & 6351 $\pm$ 85 & 4.3 $\pm$ 0.13 & -0.12 $\pm$ 0.09 & 1294 $\pm$  67 \\
 9 & 5998 $\pm$ 85 & 0.7 $\pm$ 0.02 & -0.85 $\pm$ 0.09 & 3290 $\pm$ 179 \\
10 & 5899 $\pm$ 85 & 2.2 $\pm$ 0.06 & -0.03 $\pm$ 0.09 & 1930 $\pm$ 101 \\
11 & 6251 $\pm$ 85 & 2.0 $\pm$ 0.06 &  0.13 $\pm$ 0.09 & 2360 $\pm$ 101
\enddata
\end{deluxetable*}

\bibliographystyle{apj.bst}
\bibliography{astero}

\end{document}